\date{September 28th 1999}
\documentstyle[11pt]{article}
\input epsf
\begin{document}
\newcommand{\Limits}
{\vspace*{-1mm}{\!\lim_{\small\begin{array}{c} N\!\!\to\!\infty \\[-1mm] n\!\to\! 0\end{array}}\!}}
\newcommand{\trace}{~{\large\rm Trace}~}
\newcommand{\s}{\varsigma}
\newcommand{\f}{\varphi}
\newcommand{\room}{\rule[-0.3cm]{0cm}{0.8cm}}
\newcommand{\smallroom}{\rule[-0.2cm]{0cm}{0.7cm}}
\newcommand{\hsp}{\hspace*{3mm}}
\newcommand{\vsp}{\vspace*{3mm}}
\newcommand{\be}{\begin{equation}}
\newcommand{\ee}{\end{equation}}
\newcommand{\bd}{\begin{displaymath}}
\newcommand{\ed}{\end{displaymath}}
\newcommand{\bdm}{\begin{displaymath}}
\newcommand{\edm}{\end{displaymath}}
\newcommand{\bea}{\begin{eqnarray}}
\newcommand{\eea}{\end{eqnarray}}
\newcommand{\sgn}{~{\rm sgn}}
\newcommand{\extr}{~{\rm extr}}
\newcommand{\Equiv}{\Longleftrightarrow}
\newcommand{\pprime}{{\prime\prime}}
\newcommand{\ppprime}{{\prime\prime\prime}}
\newcommand{\notexists}{\exists\hspace*{-2mm}/}
\newcommand{\bra}{\langle}
\newcommand{\ket}{\rangle}
\newcommand{\bigbra}{\left\langle\room}
\newcommand{\bigket}{\right\rangle\room}
\newcommand{\bras}{\langle\!\langle}
\newcommand{\kets}{\rangle\!\rangle}
\newcommand{\bigbras}{\left\langle\!\!\!\left\langle\room}
\newcommand{\bigkets}{\room\right\rangle\!\!\!\right\rangle}
\newcommand{\order}{{\cal O}}
\newcommand{\minus}{\!-\!}
\newcommand{\plus}{\!+\!}
\newcommand{\erf}{{\rm erf}}
\newcommand{\bbf}{{\mbox{\boldmath $f$}}}
\newcommand{\bk}{\mbox{\boldmath $k$}}
\newcommand{\bm}{\mbox{\boldmath $m$}}
\newcommand{\br}{\mbox{\boldmath $r$}}
\newcommand{\bq}{\mbox{\boldmath $q$}}
\newcommand{\bu}{\mbox{\boldmath $u$}}
\newcommand{\bx}{\mbox{\boldmath $x$}}
\newcommand{\bz}{\mbox{\boldmath $z$}}
\newcommand{\bA}{\mbox{\boldmath $A$}}
\newcommand{\bB}{\mbox{\boldmath $B$}}
\newcommand{\bC}{\mbox{\boldmath $C$}}
\newcommand{\bF}{\mbox{\boldmath $F$}}
\newcommand{\bH}{\mbox{\boldmath $H$}}
\newcommand{\bJ}{\mbox{\boldmath $J$}}
\newcommand{\bM}{\mbox{\boldmath $M$}}
\newcommand{\bQ}{\mbox{\boldmath $Q$}}
\newcommand{\bR}{\mbox{\boldmath $R$}}
\newcommand{\bW}{\mbox{\boldmath $W$}}
\newcommand{\hmu}{\hat{\mu}}
\newcommand{\hf}{\hat{f}}
\newcommand{\hQ}{\hat{Q}}
\newcommand{\hR}{\hat{R}}
\newcommand{\hbf}{\hat{\mbox{\boldmath $f$}}}
\newcommand{\hbh}{\hat{\mbox{\boldmath $h$}}}
\newcommand{\hbm}{\hat{\mbox{\boldmath $m$}}}
\newcommand{\hbr}{\hat{\mbox{\boldmath $r$}}}
\newcommand{\hbq}{\hat{\mbox{\boldmath $q$}}}
\newcommand{\hbD}{\hat{\mbox{\boldmath $D$}}}
\newcommand{\hbJ}{\hat{\mbox{\boldmath $J$}}}
\newcommand{\hbQ}{\hat{\mbox{\boldmath $Q$}}}
\newcommand{\hbR}{\hat{\mbox{\boldmath $R$}}}
\newcommand{\hbW}{\hat{\mbox{\boldmath $W$}}}
\newcommand{\bsigma}{\mbox{\boldmath $\sigma$}}
\newcommand{\btau}{{\mbox{\boldmath $\tau$}}}
\newcommand{\bomega}{{\mbox{\boldmath $\Omega$}}}
\newcommand{\bOmega}{{\mbox{\boldmath $\Omega$}}}
\newcommand{\bDelta}{{\mbox{\boldmath $\Delta$}}}
\newcommand{\bphi}{{\mbox{\boldmath $\Phi$}}}
\newcommand{\bpsi}{{\mbox{\boldmath $\psi$}}}
\newcommand{\bdelta}{{\mbox{\boldmath $\Delta$}}}
\newcommand{\btheta}{{\mbox{\boldmath $\theta$}}}
\newcommand{\bxi}{{\mbox{\boldmath $\xi$}}}
\newcommand{\bmu}{{\mbox{\boldmath $\mu$}}}
\newcommand{\brho}{{\mbox{\boldmath $\rho$}}}
\newcommand{\bEta}{{\mbox{\boldmath $\eta$}}}
\newcommand{\G}{{\cal G}}
\newcommand{\A}{{\cal A}}
\newcommand{\B}{{\cal B}}
\newcommand{\C}{{\cal C}}
\newcommand{\K}{{\cal K}}
\newcommand{\cA}{{\cal A}}
\newcommand{\cB}{{\cal B}}
\newcommand{\cC}{{\cal C}}
\newcommand{\cD}{{\cal D}}
\newcommand{\cE}{{\cal E}}
\newcommand{\cF}{{\cal F}}
\newcommand{\cL}{{\cal L}}
\newcommand{\unity}{{\bf 1}\hspace{-1mm}{\bf I}}
\newcommand{\inn}{\!\cdot\!}
\newcommand{\set}{{\tilde{D}}}
\newcommand{\sets}{{\rm sets}}
\newcommand{\ketset}{\ket_{\!\tilde{D}}}
\newcommand{\all}{D}
\newcommand{\ketall}{\ket_{\!D}}
\newcommand{\LG}{\mbox{\normalsize ${\cal G}$}}
\newcommand{\LPsi}{\mbox{\normalsize $\Psi$}}
\newcommand{\hatildeG}{\bar{G}}
\newcommand{\ketop}{\mbox{\tiny ${\rm Q\!R\!P}$}}
\newcommand{\bsomega}{\!\!\mbox{\scriptsize\boldmath $\Omega$}}
\newcommand{\rmsets}{\!\!\mbox{\scriptsize\boldmath $\Xi$}}
\newcommand{\bnul}{\mbox{\boldmath $0$}}

\newcommand{\authorA}{A.C.C. Coolen \\[1mm]
Department of Mathematics\\
King's College London\\
Strand, London WC2R 2LS, UK}
\newcommand{\authorB}{D. Saad \\[1mm]
The Neural Computing Research Group \\
Aston University\\
Birmingham B4 7ET, UK}
\title{\bf Dynamics of Learning with Restricted Training Sets\\
II. Tests and Applications}
\author{\authorA\and\authorB}

\maketitle

\begin{abstract}
\noindent
We apply a
general theory describing the dynamics of supervised learning in
layered neural networks in the regime where the size $p$ of the
training set is proportional to the number of inputs $N$, as
developed in a previous paper, to several choices of learning
rules. In the case of (on-line and batch) Hebbian learning,
where a direct exact solution is possible, we show that our theory
provides exact results at any time in many different verifiable cases.
For non-Hebbian learning rules, such as Perceptron and AdaTron, we
find very good agreement between the predictions of our theory and 
numerical simulations. Finally, we derive
three approximation schemes aimed at eliminating the need
to solve a functional saddle-point equation at each
time step, and assess their performance.
The simplest of these schemes leads to a fully
explicit and relatively simple non-linear diffusion equation for the joint field
distribution, which already
describes the learning dynamics surprisingly well over a wide range of parameters.
\end{abstract}

{\small 
\vspace*{1mm}

\begin{center}{PACS: 87.10.+e, 02.50.-r, 05.20.-y}\end{center}

\vspace*{-8mm}

\tableofcontents }

\clearpage

\section{Introduction}

In a previous paper \cite{CoolenSaad1} we have applied the formalism of dynamical
replica theory \cite{Coolenetal} to analyse the dynamics of supervised
learning in perceptrons with restricted training sets.
For an introduction into the area of the dynamics of learning
in layered neural networks, a guide to the relevant references, as well as a
proper discussion of the peculiarities of the dynamics of learning with restricted
as opposed to infinite raining sets, we refer to \cite{CoolenSaad1}.
The microscopic variables in the learning process are the
components of the weight vector $\bJ$ of the `student' network. The `teacher' network, which defines the
task to be learned by the student, is characterised by a weight
vector $\bB$. Learning proceeds on the basis of answers to be given to questions
$\bxi\in\set\subseteq\{-1,1\}^N$, according to a dynamical rule for $\bJ$ which is defined in terms of a
function $\G[x,y]$, where $x=\bJ\cdot\bxi$ and $y=\bB\cdot\bxi$ are the
student and teacher fields, respectively. 
 The randomly composed set $\set$ of questions,
of size $p=\alpha N$, is called the training set. 
If $\alpha<\infty$ as $N\to\infty$ the learning dynamics will
be nontrivial. Firstly, the data in $\set$ will be recycled in due
course, which generates complicated correlations
and non-Gaussian local field distributions, and allows
the system to improve its performance partly by memorizing
answers rather than by learning the underlying rule (hence the
difference between training- and generalization errors).
Secondly,  the actual
composition of the randomly drawn training set introduces an element of frozen disorder into the
problem, which will have to be averaged out.
The analysis described in \cite{CoolenSaad1} resulted
in a general macroscopic theory, describing the behaviour of
such learning processes with $\alpha<\infty$ in the limit $N\to\infty$ (infinite systems) and on finite
time-scales, in terms of deterministic laws for macroscopic observables.
The theory applies to both on-line learning (where weight updates are made
after each presentation of an input vector from the training set),
and to batch learning (where weight updates are themselves
averages over the full training set), as well as to arbitrary
learning rules (i.e. arbitrary functions $\G[x,y]$).

In this paper we apply this general theory to various different specific choices of
learning rules. One of these, (on-line and batch) Hebbian
learning, provides an excellent benchmark test for our theory, since
for this simple rule exact
solutions are known, even for the regime of restricted training sets
\cite{Raeetal}. We find that our theory is fully exact for batch
execution, and that it succeeds in predicting exactly the evolution of
several macroscopic observables, including the
generalisation error and moments of the joint field distribution
for student and teacher fields, in the on-line case (although here full exactness is
difficult to assess, and not  a priori guaranteed).
A preliminary presentation of some of the results in this paper
(those involving Hebbian learning) was given in
\cite{CoolenSaadNewton}.  
For non-Hebbian error-correcting learning rules, such as on-line and
batch versions of Perceptron learning  and AdaTron learning, no exact
solutions are known at present with which to confront our theory; 
instead we here compare the
predictions (with regard to the evolution of training- and generalization errors and the joint field
distribution)
of the full theory, as well as of
a number of simple approximations of our equations,  with the
results of carrying out
extensive  numerical simulations in large (size $N=10,000$) neural networks.
We find, surprisingly, that
even the simplest of these approximations, which does not require
solving any saddle point equations and takes the form of a
fully explicit non-linear diffusion
equation for the joint field distributions $P[x,y]$, describes the simulation
experiments remarkably well. Employing the more sophisticated (and
thereby more CPU intensive) approximations, or, at the other end of the spectrum,
a numerical solution of the full macroscopic theory, leads to increasingly accurate quantitative
predictions for the evolution of the relevant macroscopic
observables of the learning process, with deviations between
theory and numerical experiment which are of the order of magnitude of the
finite size effects in the simulations.  
We close our paper with a
discussion of the strengths and weaknesses of the approach used,
and an outlook on future work on the dynamics of learning with
restricted training sets, involving the present and possibly other formalisms.

\section{Summary of the Theory and its Properties}

In this section we will be brief in order to avoid inappropriate duplication of the
material in \cite{CoolenSaad1}, to which we  refer
for full details.

\subsection{The Macroscopic Laws}

Our macroscopic observables are   $Q=\bJ^2$,
$R=\bB\inn\bJ$,
and the joint distribution of student and teacher fields (or
`activations')
$P[x,y]=\bra\delta[x\minus \bJ\inn\bxi]\delta[y\minus
\bB\inn\bxi]\ketset$. For $N\to\infty$ all these quantities are found to obey
deterministic and self-averaging
equations. 
We define $\bra f[x,y]\ket =\int\!dx Dy ~P[x|y]f[x,y]$, 
where $Dy=(2\pi)^{-{\frac{1}{2}}}e^{-\frac{1}{2}y^2}dy$, 
and the
following averages
(the function $\Phi[x,y]$ will be specified below):
\be
U=\bra \Phi[x,y]\G[x,y]\ket~~~~~~~~~
V=\bra x \G[x,y]\ket~~~~~~~~~
W=\bra y \G[x,y]\ket~~~~~~~~~
Z=\bra \G^2[x,y]\ket
\label{eq:fouraverages}
\ee
For on-line learning (which is a stochastic process) we have found,
using the prescriptions of dynamical replica theory (in replica
symmetric ansatz)
\cite{CoolenSaad1}:
\be
\frac{d}{dt}Q=2\eta V +\eta^2 Z
~~~~~~~~~~~~~~~~
\frac{d}{dt}R =
\eta W
\label{eq:summaryonlinedQdtdRdt}
\ee
\bd
\frac{d}{dt}P[x|y]=\frac{1}{\alpha}\int\!dx^\prime P[x^\prime|y]\left[
\delta[x\minus
x^\prime\minus \eta\G[x^\prime\!,y]]\minus \delta[x\minus
x^\prime]\right]
- \eta\frac{\partial}{\partial x} \left\{\room
P[x|y]\left[ U(x\minus Ry) \plus Wy\right]\right\}
\ed
\be
+ \frac{1}{2}\eta^2 Z
\frac{\partial^2}{\partial x^2}P[x|y]
- \eta\left[V\! \minus RW\! \minus (Q\minus R^2)U \right]
\frac{\partial}{\partial x}\left\{\room
P[x|y]\Phi[x,y]
\right\}
\label{eq:summaryonlinedPdt}
\ee
For batch learning (which is a deterministic process) we found:
\be
\frac{d}{dt}Q=
2\eta V
~~~~~~~~~~~~~~~
\frac{d}{dt}R
=\eta W
\label{eq:summarybatchdQdtdRdt}
\ee
\bd
\frac{d}{dt}P[x|y]=
- \frac{\eta}{\alpha}\frac{\partial}{\partial x}\left[
P[x|y]\G[x,y]\right]
- \eta\frac{\partial}{\partial x} \left\{\room
P[x|y]\left[ U(x\minus Ry) \plus Wy\right]\right\}
\ed
\be
- \eta\left[V\! \minus RW\! \minus (Q\minus R^2)U \right]
\frac{\partial}{\partial x}\left\{\room
P[x|y]\Phi[x,y]
\right\}
\label{eq:summarybatchdPdt}
\ee
From the solution of the above closed sets of equations for the trio $\{Q,R,P\}$ (one of which is a function)
follow the familiar training- and generalization
errors  $E_{\rm t}=\bra \theta[-xy]\ket$
and $E_{\rm g}=\pi^{-1}\arccos[R/\sqrt{Q}]$.
The auxiliary order parameters generated in the replica calculation,
the spin-glass order parameter $q$ and the function
$M[x|y]$, are calculated at each time-step by solving the following saddle-point
equations:
\be
\bra (x\minus Ry)^2\ket
+(qQ\minus R^2)(1\minus \frac{1}{\alpha})
=\left[2(qQ\minus R^2)^\frac{1}{2}\plus \frac{1}{B}\right]
\int\!DyDz~ z\bra x\ket_\star
\label{eq:summarysaddle_q}
\ee
\be
P[X|y]
=\int\! Dz~ \bra \delta[X\minus x]\ket_\star
\label{eq:summarysaddle_M}
\ee
with
\be
B=\frac{\sqrt{qQ\minus R^2}}{Q(1\minus q)}
~~~~~~~~~~~~~
\bra f[x,y,z]\ket_\star =
 \frac{\int\!dx~M[x|y]e^{Bxz}f[x,y,z]}
{\int\!dx~M[x|y]e^{Bxz}}
\label{eq:notation}
\ee
Without loss of generality we can always normalize $M$ according to $\int\!dx~M[x|y]=1$ for all $y\in\Re$.
From the physical meaning of $q$ follows
$R^2/Q \leq q\leq 1$.
After $q$ and $M[x,y]$ have been determined,
the key function $\Phi[x,y]$ in
(\ref{eq:fouraverages},\ref{eq:summaryonlinedPdt},\ref{eq:summarybatchdPdt})
is calculated as
\be
\Phi[X,y]=
\left\{\room Q(1\minus q)P[X|y]\right\}^{-1}\!\int\!Dz~
\bra X\minus x \ket_\star
\bra \delta[X\minus x]\ket_\star
\label{eq:defPhi}
\ee
We refer to \cite{CoolenSaad1} for full derivations  of the above
equations.

\subsection{Properties of the Macroscopic Laws}

Some useful properties of our theory are independent of which learning
rule $\G[x,y]$ is used. The first of these is that in the limit
$\alpha\to\infty$, 
which corresponds to the case of 
infinite training sets, our theory reduces to the simpler formalism
initiated in \cite{KinzelRujan} and elaborated in papers  like
\cite{BiehlSchwarze,KinouchiCaticha}, which is built on  assuming $P[x,y]$
to be of a Gaussian form (this only happens as
$\alpha\to\infty$) \cite{CoolenSaad1}.
Secondly, for any given $q$ the solution $M[x|y]$ of the functional
saddle-point problem (\ref{eq:summarysaddle_M}) is unique, and can
even be obtained as the fixed-point of a converging nonlinear
functional map \cite{CoolenSaad1}.
Thirdly, we note that the first conditional
moment $\overline{x}(y)=\int\!dx~xP[x|y]$ of $P[x|y]$ of the joint field distribution obeys a
simple equation, which is obtained from
(\ref{eq:summaryonlinedPdt}) and (\ref{eq:summarybatchdPdt}) upon
multiplication by $x$, followed by integration over $x$:
\be
\frac{d}{dt}[\overline{x}(y)-Ry]=
\frac{\eta}{\alpha}\int\!dx~P[x|y] \G[x,y]
+\eta  U[\overline{x}(y)\minus Ry]
\label{eq:moment}
\ee
where we have also used the built-in property
$\int\!dx~P[x|y]\Phi[x,y]=0$ for all $y$.
\vsp

Alternatively we could rewrite our macroscopic equations into Fourier language,
i.e. in terms of $\hat{P}[k|y]=\int\!dx~e^{-ikx}P[x|y]$ and
$\hat{M}[k|y]=\int\!dx~e^{-ikx}M[x|y]$.
The functional saddle-point equation, giving $\hat{M}[k]$, then  becomes
\be
\hat{P}[k|y]=\int\!Dz ~\frac{\hat{M}[k\plus iBz|y]}{\hat{M}[iBz|y]}
\label{eq:FourierSPE}
\ee
and the diffusion equation takes the form
\bd
\frac{d}{dt}\log\hat{P}[k|y]~=~
\frac{1}{\alpha}
\left\{\int\!dk^\prime~\frac{\hat{P}[k^\prime|y]}{\hat{P}[k|y]}
\int\!\frac{dx^\prime}{2\pi}e^{ix^\prime(k^\prime-k)
-i\eta k\G[x^\prime,y]}
-1\right\}
- i\eta k (W\!\minus UR)y
\ed
\be
+~\eta k U\frac{\partial}{\partial k}\log\hat{P}[k|y]
- \frac{1}{2}\eta^2 k^2 Z
~-~ i\eta k
\left[\frac{V\!\minus RW\!\minus (Q\minus R^2)U}{\sqrt{qQ\minus R^2}\hat{P}[k|y]}\right]
\int\!Dz ~z~\frac{\hat{M}[k+iBz|y]}{\hat{M}[iBz]}
\label{eq:FourierdPdt}
\ee
This representation will be particularly convenient when applying our theory to
Hebbian learning rules.
We can derive more explicit results for the
special class of locally Gaussian solutions, defined as
\bd
P[x|y]=\frac{e^{-\frac{1}{2}[x-\overline{x}(y)]^2/\Delta^2(y)}}{\Delta(y)\sqrt{2\pi}}
\ed
For such distributions the functional saddle-point equation (\ref{eq:summarysaddle_M})
can be solved, giving
\be
M[x,y]=
\frac{e^{-\frac{1}{2}[x-\overline{x}(y)]^2/\sigma^2(y)}}{\sigma(y)\sqrt{2\pi}}
\label{eq:localgauss1}
\ee
with
$\Delta^2(y)=\sigma^2(y)\plus B^2\sigma^4(y)$. For such solutions to exist,
the conditional moments $\overline{x}(y)$ and $\Delta(y)$ must
obey the following equation:
\bd
-ik\frac{d}{dt}\overline{x}(y)-\frac{1}{2}k^2\frac{d}{dt}\Delta^2(y)
~=~
\frac{1}{\alpha}\left\{
\int\!\frac{du}{\sqrt{2\pi}}
e^{-\frac{1}{2}[u-ik\Delta(y)]^2-ik\eta \G[\overline{x}(y)+u\Delta(y),y]}
-1\right\}
- i\eta k\left\{ Wy +U[\overline{x}(y)\minus Ry]\right\}
\ed
\be
- \frac{1}{2}k^2 \left\{\eta^2  Z +2 \eta U\Delta^2(y)
~+ 2\eta\sigma^2(y)
\left[\frac{V\!\minus RW\!\minus (Q\minus R^2)U}
{Q(1\minus q)}\right]\right\}
\label{eq:FourierGaussiandPdt}
\ee
The simple form of (\ref{eq:localgauss1}) allows us to calculate many objects explicitly. In particular:
\bd
\bra x\ket_\star = \overline{x}(y)+zB\sigma^2(y)
~~~~~~~~~~~~~~~~~~
\Phi[x,y]=
\frac{x-\overline{x}(y)}{Q(1\minus q)[1\plus B^2\sigma^2(y)]}
\ed
\vsp

Finally, for many types of learning rules there are symmetry properties of our macroscopic  equations to be
exploited.
Note that in the most common types of learning rules no distinction is being
made between system errors of the type $\{x>0,~y<0\}$ and those where
$\{x<0,~y>0\}$. This translates into the following property of the function
$\G[x,y]$ (note that by the nature of the learning processes under
study $\G[x,y]$ can depend on $y$ only via $\sgn(y)$):
\be
\G[x,y]=\sgn(y)\left\{\room \theta[xy] \G_+[|x|]+\theta[-xy]\G_-[|x|]\right\}
\label{eq:specialform}
\ee
In particular we find for the most common learning recipes:
\bd
\begin{array}{lllll}
{\rm Hebbian:}    && \G_+[u]=1, && \G_-[u]=1\\[1mm]
{\rm Perceptron:} && \G_+[u]=0, && \G_-[u]=1\\[1mm]
{\rm AdaTron:}    && \G_+[u]=0, && \G_-[u]=u\\
\end{array}
\ed
The form (\ref{eq:specialform}) implies the symmetry $\G[-x,y]=\G[x,- y]$, which turns out to allow for self-consistent solutions of the
macroscopic equations with the following property at any time:
\be
P[- x|y]=P[x|- y]
\label{eq:symmetry_P}
\ee
Combination of (\ref{eq:symmetry_P}) with
(\ref{eq:summarysaddle_M})
and (\ref{eq:defPhi})
shows that the measure $M[x|y]$ and the function $\Phi[x,y]$ must
consequently have the following symmetry  properties:
\bd
M[- x|y]=M[x|- y],~~~~~~~~~~
\Phi[- x,y]=-\Phi[x,- y]
\ed
This, in turn, guarantees that the right-hand sides of equations
(\ref{eq:summaryonlinedPdt}) and (\ref{eq:summarybatchdPdt}) indeed
preserve the symmetry (\ref{eq:symmetry_P}) of the field distribution,
as claimed.

\clearpage
\section{Benchmark Tests: Hebbian Learning}

In the special case of the Hebb rule, $\G[x,y]\!=\!\sgn[y]$, where
weight changes $\Delta \bJ$ never depend on  $\bJ$, one can
write down an explicit expression for the weight
vector $\bJ$ at any time, and thus for the expectation values of our
observables. We choose as our initial field distribution a simple Gaussian one,
resulting from an initialization process which did not involve
the training set:
\be
P_0[x|y]=\frac{e^{-\frac{1}{2}(x-R_0y)^2/(Q_0-R_0^2)}}{\sqrt{2\pi(Q_0\minus R_0^2)}}
\label{eq:Pinitial}
\ee
Careful averaging of the exact expressions for our observables
over all `paths' $\{\bxi(0),\bxi(1),\ldots\}$ taken by the
question/example 
vector through the training set $\set$ (for on-line learning),
followed by averaging over all
realizations of the
training set $\set$ of size $p=\alpha N$, and taking the
$N\to\infty$ limit,
then leads to the
following {\em exact} result \cite{Raeetal}.
For on-line Hebbian
learning one ends up with:
\be
Q=Q_0+2\eta t R_0 \sqrt{\frac{2}{\pi}} +\eta^2 t +\eta^2 t^2
\left[\frac{1}{\alpha}\plus \frac{2}{\pi}\right]
~~~~~~~~
R=R_0 +\eta t\sqrt{\frac{2}{\pi}}
\label{eq:QRexactonline}
\ee
\be
P[x|y]=\int\!\frac{d\hat{x}}{2\pi} ~e^{-\frac{1}{2}\hat{x}^2
[Q-R^2]+i\hat{x}[x-Ry]+\frac{t}{\alpha}[e^{-i\eta \hat{x}\sgn[y]}-1]}
\label{eq:Pexactonline}
\ee
For batch learning a similar calculation\footnote{Note that in \cite{Raeetal} only
the on-line calculation was carried out; the batch calculation
can be done along the same lines.} gives:
\be
Q=Q_0+2\eta t R_0 \sqrt{\frac{2}{\pi}} +\eta^2 t^2
\left[\frac{1}{\alpha}\plus \frac{2}{\pi}\right]
~~~~~~~~~~
R=R_0 +\eta t\sqrt{\frac{2}{\pi}}
\label{eq:QRexactbatch}
\ee
\be
P[x|y]=\frac{e^{-\frac{1}{2}[x-Ry-(\eta
t/\alpha)\sgn[y]]^2/(Q-R^2)}}{\sqrt{2\pi(Q\minus R^2)}}
\label{eq:Pexactbatch}
\ee
Neither of the two field distributions is of a fully Gaussian
form (although the batch distribution is at least conditionally Gaussian).
Note that for both on-line and batch Hebbian learning we have
\be
\int\!dx~x P[x|y]=Ry+\frac{\eta t}{\alpha}\sgn[y]
\label{eq:Hebbxaverage}
\ee
The generalization- and
training errors are, as before, given in terms of the above observables as $E_{\rm
g}\!=\!\pi^{-1}\arccos[R/\sqrt{Q}]$ and $E_{\rm t}\!=\!\int\!Dy dx
P[x|y]\theta[- xy]$.
We thus have exact expressions for both the generalization error and the
training error at any time and for any $\alpha$.
The asymptotic values, for both batch and on-line Hebbian learning,
are given by
\be
\lim_{t\to\infty}E_{\rm g}=\frac{1}{\pi}\arccos
\left[\frac{1}{\sqrt{1+\pi/2\alpha}}\right]
\label{eq:Egstat}
\ee
\be
\lim_{t\to\infty}
E_{\rm t}=\frac{1}{2}-\frac{1}{2}\int\!Dy~
{\rm erf}\left[|y|\sqrt{\frac{\alpha}{\pi}}\plus
\frac{1}{\sqrt{2\alpha}}\right]
\label{eq:Etstat}
\ee
As far as $E_{\rm g}$ and $E_{\rm t}$ are concerned, the
differences between batch and on-line Hebbian learning are
confined to transients.
Clearly, the above exact results (which can only be obtained for Hebbian-type
learning rules) provide excellent and welcome benchmarks with which to test
general theories such as the one investigated in the present paper.

\subsection{Batch Hebbian Learning}

We compare the exact solutions for Hebbian learning
to the predictions of our general
theory, turning first to batch Hebbian learning.
We insert into the equations of our general formalism the
Hebbian recipe $\G[x,y]\!=\!\sgn[y]$. This simplifies
our dynamic equations enormously.
In particular we obtain:
\bd
U=0,~~~~~~~~~~~~
V=\bra x \sgn(y)\ket,~~~~~~~~~~~~
W=\sqrt{2/\pi}
\ed
For batch learning we consequently find:
\bd
\frac{d}{dt}Q=2\eta V
~~~~~~~~~~~~~~~~
\frac{d}{dt}R =
\eta \sqrt{2/\pi}
\ed
\bd
\frac{d}{dt}P[x|y]=-\frac{\eta}{\alpha}\sgn(y)\frac{\partial}{\partial x}P[x|y]
- \eta y\sqrt{\frac{2}{\pi}} \frac{\partial}{\partial x} P[x|y]
- \eta(V\! \minus R\sqrt{\frac{2}{\pi}})
\frac{\partial}{\partial x}\left\{\room
P[x|y]\Phi[x,y]
\right\}
\ed
Given the initial field distribution  (\ref{eq:Pinitial}),  we
immediate obtain 
$V_0\!=\!R_0\sqrt{2/\pi}$.
From the general property $\int\!dx~P[x|y]\Phi[x,y]\!=\!0$ and the
above diffusion equation for $P[x|y]$ we derive an equation for
the quantity $V\!=\!\bra x\sgn(y)\ket$,
resulting in
$\frac{d}{dt}V=\eta/\alpha +2\eta/\pi$, which subsequently
allows us to solve
\be
Q=
Q_0+2\eta t R_0 \sqrt{\frac{2}{\pi}} +\eta^2 t^2
\left[\frac{1}{\alpha}\plus \frac{2}{\pi}\right]
~~~~~~~~~~~~~~~~
R=R_0 +\eta t\sqrt{\frac{2}{\pi}}
\label{eq:QRDRTbatch}
\ee
Furthermore, it turns out that the above diffusion equation for
$P[x|y]$ meets the requirements for having conditionally-Gaussian solutions, i.e.
\bd
P[x|y]=\frac{e^{-\frac{1}{2}[x-\overline{x}(y)]^2/\Delta^2(y)}}{\Delta(y)\sqrt{2\pi}},
~~~~~~~~~~~~
M[x|y]=
\frac{e^{-\frac{1}{2}[x-\overline{x}(y)]^2/\sigma^2(y)}}{\sigma(y)\sqrt{2\pi}}
\ed
provided the $y$-dependent average $\overline{x}(y)$ and the
$y$-dependent variances $\Delta(y)$ and $\sigma(y)$
obey the following three coupled equations:
\bd
\overline{x}(y)=Ry\plus\frac{\eta t}{\alpha}\sgn(y)
~~~~~~~~~~~~
\frac{d}{dt}
\Delta^2(y)=\frac{2\eta^2 t \sigma^2(y)}{\alpha Q(1\minus q)}
~~~~~~~~~~~~
\Delta^2(y)=\sigma^2(y)\plus B^2\sigma^4(y)
\ed
The spin-glass order parameter $q$ is to be solved from the remaining scalar saddle-point
equation (\ref{eq:summarysaddle_q}). With help of identities like
$\bra x\ket_\star = \overline{x}(y)+zB\sigma^2(y)$, which only
hold
for conditionally-Gaussian solutions,  one can simplify
the latter to
\bd
\frac{\eta^2 t^2}{\alpha}+\alpha\int\!Dy~\Delta^2(y)
+(qQ\minus R^2)(\alpha \minus 1)
=\alpha \left[2\frac{qQ\minus R^2}{Q(1\minus q)}\plus 1\right]
\int\!Dy~ \sigma^2(y)
\ed
We now immediately find the solution
\bd
\Delta^2(y)=Q\minus R^2,~~~~~~~~~
\sigma^2(y)=Q(1\minus q),~~~~~~~~~
q=[\alpha R^2\plus \eta^2t^2]/\alpha Q
\ed
\be
P[x|y]=\frac{e^{-\frac{1}{2}[x-Ry-(\eta
t/\alpha)\sgn(y)]^2/(Q-R^2)}}{\sqrt{2\pi(Q\minus R^2)}}
\label{eq:PDRTbatch}
\ee
(this solution is unique).
If we calculate the generalization error
and the training error from
(\ref{eq:QRDRTbatch}) and (\ref{eq:PDRTbatch}), respectively, we
recover  the
exact expressions
\be
E_{\rm g}=\frac{1}{\pi}\arccos\left[\frac{R_0 \plus \eta
t\sqrt{\frac{2}{\pi}}}
{\sqrt{Q_0\plus 2\eta t R_0 \sqrt{\frac{2}{\pi}} \plus \eta^2 t^2
\left[\frac{1}{\alpha}\plus \frac{2}{\pi}\right]}}\right]
\label{eq:EgDRTbatch}
\ee
\be
E_{\rm t}
=\frac{1}{2}-\frac{1}{2}\int\!Dy~
{\rm erf}\left[\frac{|y|[R_0\plus \eta t\sqrt{\frac{2}{\pi}}] \plus
\frac{\eta t}{\alpha}}{\sqrt{2[Q_0\minus R_0^2\plus \frac{\eta^2 t^2}{\alpha}]}}\right]
\label{eq:EtDRTbatch}
\ee
Comparison of
(\ref{eq:QRDRTbatch},\ref{eq:PDRTbatch}) with
(\ref{eq:QRexactbatch},\ref{eq:Pexactbatch})
shows that for batch Hebbian learning our theory is fully exact.
This is not a big feat as far as $Q$ and $R$ (and thus $E_{\rm g}$)
are concerned, whose determination  did
not require knowing the function $\Phi[x,y]$. The fact that our theory
also gives the exact values for $P[x|y]$ and $E_{\rm t}$, however, is less
trivial,
since here the disordered nature of the
learning dynamics, leading to non-Gaussian distributions, is truly
relevant.

\subsection{On-Line Hebbian Learning}

We next insert the
Hebbian recipe $\G[x,y]\!=\!\sgn[y]$ into the on-line equations
(\ref{eq:summaryonlinedQdtdRdt},\ref{eq:summaryonlinedPdt}).
Direct analytical
solution of these equations,
or a demonstration
that they are solved by the exact result
(\ref{eq:QRexactonline},\ref{eq:Pexactonline}),
although not ruled out, has not yet
been achieved by us.
 The reason is that here one has conditionally Gaussian field distributions
only in special
limits. Numerical solution is in principle straightforward,
but will be quite CPU intensive (see also a subsequent section).
For  small learning rates the on-line equations reduce to the
batch ones, so we know that in first order in $\eta$ our
on-line equations are exact (for any $\alpha$, $t$).
We now show that
the predictions of our theory are fully exact (i) for $Q$, $R$ and
$E_{\rm g}$, (ii) for the first moment (\ref{eq:Hebbxaverage})
of the conditional field distribution, and
(iii) for all order parameters in the stationary state. At
intermediate times we construct an approximate solution of
our equations in order to obtain predictions for $P[x|y]$
and $E_{\rm t}$.

As before we choose a Gaussian initial field distribution.
Many (but not all) of our previous simplifications still hold, e.g.
\bd
U=0,~~~~~~~~~~~
V=\bra x \sgn(y)\ket,~~~~~~~~~~~
W=\sqrt{2/\pi},~~~~~~~~~~~
Z=1
\ed
($Z$ did not occur in the
batch equations).
Thus for on-line learning we find:
\bd
\frac{d}{dt}Q=2\eta V +\eta^2
~~~~~~~~~~~~~~~~~~~~~
\frac{d}{dt}R =
\eta \sqrt{2/\pi}
\ed
The previous derivation of the identities
$\frac{d}{dt}V\!=\!\eta/\alpha \plus 2\eta/\pi$ and
$V_0=R_0\sqrt{2/\pi}$
still applies (just replace the batch
diffusion equation by the
on-line one),
but
the resultant expression  for $Q$ is different.
Here we obtain:
\be
Q=Q_0+2\eta t R_0 \sqrt{\frac{2}{\pi}} +\eta^2 t
+ \eta^2 t^2
\left[\frac{1}{\alpha}\plus \frac{2}{\pi}\right]
~~~~~~~~~~~~~
R=R_0 +\eta t\sqrt{\frac{2}{\pi}}
\label{eq:QRDRTonline}
\ee
Comparing (\ref{eq:QRDRTonline}) with (\ref{eq:QRexactonline})
reveals that also for on-line Hebbian learning our
theory is exact with regard to $Q$ and $R$, and
thus also with regard to $E_{\rm g}$.
Upon using $V\!=\!\eta t/\alpha \plus R\sqrt{2/\pi}$, the
on-line diffusion equation simplifies to
\bd
\frac{d}{dt}P[x|y]=\frac{1}{\alpha}\left\{\room
P[x\minus \eta\sgn(y)|y]\minus P[x|y]\right\}
- \eta y\sqrt{\frac{2}{\pi}} \frac{\partial}{\partial x} P[x|y]
+ \frac{1}{2}\eta^2
\frac{\partial^2}{\partial x^2}P[x|y]
- \frac{\eta^2 t}{\alpha}
\frac{\partial}{\partial x}\left\{\room
P[x|y]\Phi[x,y]
\right\}
\ed
Multiplication of this equation by $x$ followed by integration over
$x$, together with usage of the general properties $\int\!dx~\{
P[x|y]\Phi[x,y]\}=0$ and
$\int\!dx~xP_0[x|y]=R_0y$, gives us the average of the
conditional distribution $P[x|y]$ at any time:
\bd
\overline{x}(y)=\int\!dx~x P[x|y]=Ry +\frac{\eta t}{\alpha}\sgn[y]
\ed
Comparison with (\ref{eq:Hebbxaverage}) shows also this prediction
to be correct.

We now turn to observables which involve
more detailed knowledge of the function $\Phi[x,y]$.
Our result for $\overline{x}(y)$ and
the identity $\bra
x\ket_\star=B^{-1}\frac{\partial}{\partial z}\log \hat{M}[iBz|y]$
allow us to rewrite all remaining equations
in Fourier representation,
i.e. in terms of $\hat{P}[k|y]\!=\!\int\!dx~e^{-ikx}P[x|y]$
and $\hat{M}[k|y]\!=\!\int\!dx~e^{-ikx}M[x|y]$:
\be
\frac{d}{dt}\log\hat{P}[k|y]=
\frac{1}{\alpha}
\left[e^{-i\eta k\sgn(y)}\minus 1\right]
- i\eta ky \sqrt{\frac{2}{\pi}}
- \frac{1}{2}\eta^2 k^2
- \frac{ik\eta^2  t}{\alpha
\hat{P}[k|y]\sqrt{qQ\minus R^2}}
\int\!Dz ~z~\frac{\hat{M}[k\plus iBz|y]}{\hat{M}[iBz|y]}
\label{eq:HebbonlineFourier}
\ee
with $\log \hat{P}_0[k|y]=- ikR_0y\minus \frac{1}{2}k^2 (Q_0\minus
R_0^2)$, and with the two saddle-point equations
\be
\hat{P}[k|y]=\int\!\!Dz ~\frac{\hat{M}[k\plus iBz|y]}{\hat{M}[iBz|y]}
\label{eq:Hebbonlinesaddle2}
\ee
\be
\frac{\eta^2 t^2}{\alpha^2}
+\int\!\!Dy\!\int\!\!dx~P[x|y][x\minus\overline{x}(y)]^2
+(1\minus \frac{1}{\alpha})(qQ\minus R^2)
=\left[2Q(1\minus q)\plus
\frac{1}{B^2}\right]\!
\int\!\!DyDz~\frac{\partial^2}{\partial z^2}\log \hat{M}[iBz|y]
\label{eq:Hebbonlinesaddle1}
\ee
Since the fields $x$
grow linearly in time (see our expression for
$\overline{x}(y)$) the
equations
(\ref{eq:HebbonlineFourier},\ref{eq:Hebbonlinesaddle1},\ref{eq:Hebbonlinesaddle2})
cannot have proper $t\!\to\!\infty$ limits.
To extract asymptotic properties we have to
turn to the rescaled distribution
$\hat{Q}[k|y]\!=\!\hat{P}[k/t|y]$.
We define $v(y)\!=\!(\eta/\alpha)\sgn(y)
\plus \eta y\sqrt{2/\pi}$.
Careful integration of
(\ref{eq:HebbonlineFourier}), followed by inserting $k\!\to
k/t$ and by taking the limit $t\!\to\!\infty$, produces:
\be
\log\hat{Q}_\infty[k|y]=-ikv(y)
-\frac{i\eta^2 k}{\alpha}\!\int_0^1\!du~
\lim_{t\to\infty} \frac{t}{\sqrt{qQ\minus
R^2}}\int\!Dz~z~\frac{\hat{M}[u k/t\plus iBz|y]}
{\hat{Q}_\infty[uk|y]\hat{M}[iBz|y]}
\label{eq:Qequil}
\ee
with the functional saddle-point equation
\be
\hat{Q}[k|y]=\int\!Dz ~\frac{\hat{M}[k/t\plus iBz|y]}
{\hat{M}[iBz|y]}
\label{eq:Mequil}
\ee
The rescaled asymptotic system
(\ref{eq:Qequil},\ref{eq:Mequil}) admits the solution
\bd
\hat{Q}[k|y]=e^{-ik v(y)-\frac{1}{2}k^2 \tilde{\Delta}^2},
~~~~~~~~~~~~
\hat{M}[k|y]=e^{-ik \overline{x}(y)
-\frac{1}{2} k^2 \tilde{\sigma}^2 t}
\ed
with the asymptotic values of
$B$, $\tilde{\Delta}$, $\tilde{\sigma}$ and $q$ determined by solving
the following equations:
\bd
\tilde{\Delta}=B\tilde{\sigma}^2
~~~~~~~~~~~~~
\tilde{\Delta}=\frac{\eta^2}{\alpha}\lim_{t\to\infty}\frac{t}{\sqrt{qQ\minus
R^2}}
~~~~~~~~~~~~~
B=\lim_{t\to\infty}\frac{\sqrt{qQ\minus R^2}}{Q(1\minus q)}
\ed
\bd
\eta^2/\alpha^2+\tilde{\Delta^2}+(1\minus\alpha^{-1})
\lim_{t\to\infty}(qQ\minus R^2)/t^2=
2B^2\tilde{\sigma}^2 \lim_{t\to\infty}Q(1\minus q)/t
\ed
Inspection shows that these four asymptotic equations are solved by
\bd
\lim_{t\to\infty}\tilde{\Delta}=\eta/\sqrt{\alpha},~~~~~~~~\lim_{t\to\infty}q=1
\ed
so that
\be
\lim_{t\to\infty}\hat{P}_t[k/t|y]=
e^{-ik\eta \left[\alpha^{-1}\sgn(y)+
y\sqrt{2/\pi}\right]-\frac{1}{2}\eta ^2 k^2/\alpha }
\label{eq:DRTequilP}
\ee
Comparison with (\ref{eq:QRexactonline},\ref{eq:Pexactonline}) shows
that this prediction (\ref{eq:DRTequilP}) is again exact. Thus the
same is true for the asymptotic training error.
\vsp

\begin{figure}[t]
\centering
\vspace*{91mm}
%fullhebbsimu.eps
\hbox to \hsize{\hspace*{-1mm}\includegraphics{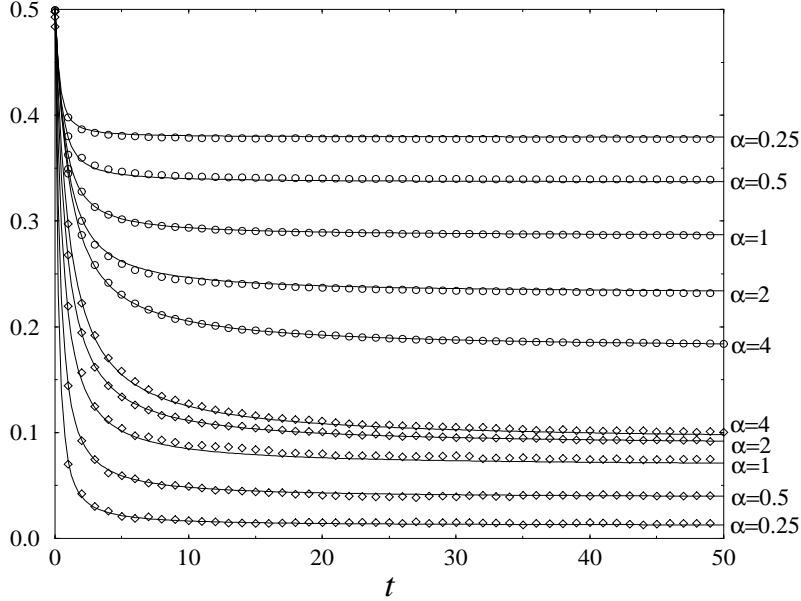}\hspace*{1mm}}
\vspace*{-10mm}
\caption{On-line Hebbian
learning, simulations versus theoretical predictions,
for $\eta=1$ and $\alpha\in\{0.25,0.5,1.0,2.0,4.0\}$ ($N=10,000$). Upper
curves: generalization errors as functions of time. Lower
curves: training errors as functions of time.
Circles: simulation results for $E_{\rm g}$;
diamonds: simulation results for $E_{\rm t}$.
Solid lines: corresponding predictions of dynamical replica theory.}
\label{fig:hebberrors}
\end{figure}

Finally, in order to arrive at predictions with respect to $P[x|y]$ and
$E_{\rm t}$ for intermediate times (without rigorous analytical
solution of the functional saddle-point equation),
and in view of the conditionally-Gaussian form of the field distribution
both at $t\!=\!0$ and at $t\!=\!\infty$, it would appear to make sense for us to  approximate
$P[x|y]$ and $M[x|y]$ by simple
conditionally Gaussian
distributions at any time:
\bd
P[x|y]=\frac{e^{-\frac{1}{2}[x-\overline{x}(y)]^2/\Delta^2}}{\Delta\sqrt{2\pi}},
~~~~~~~~~~~~~~~~
M[x|y]=
\frac{e^{-\frac{1}{2}[x-\overline{x}(y)]^2/\sigma^2}}{\sigma\sqrt{2\pi}}
\ed
with the (exact) first moments
$\overline{x}(y)=R y\plus \eta
t\alpha^{-1}\sgn(y)$,
and with the variance $\Delta^2$ self-consistently
given by the solution of:
\bd
\Delta^2=\sigma^2\plus B^2\sigma^4
~~~~~~~~~~~~~
B=\frac{\sqrt{qQ\minus R^2}}{Q(1\minus q)}
~~~~~~~~~~~~~
\frac{d}{dt}\Delta^2
=
\frac{\eta^2}{\alpha}\plus \eta^2\plus
\frac{2 \eta^2 t \sigma^2}{\alpha Q(1\minus q)}
\ed
\bd
\alpha\Delta^2+\frac{\eta^2 t^2}{\alpha}
+(qQ\minus R^2)(\alpha\minus 1)
=\alpha \sigma^2\left[2\frac{qQ\minus R^2}{Q(1\minus q)}\plus 1\right]
\ed
The solution of the above coupled equations behaves as
\bd
\begin{array}{lll}
\Delta^2 = Q- R^2 + \eta^2 t/\alpha+\order(t^3)&& (t\to 0)\\[4mm]
\Delta^2 = (Q\minus R^2)[1\plus \order(t^{-1})] && (t\to \infty)
\end{array}
\ed
for short and long times, respectively (note $Q\minus R^2\sim t^2$ as
$t\!\to\!\infty$).  Thus we obtain a simple approximate
solution of our equations, which extrapolates between exact results at
the temporal boundaries $t\!=\!0$ and $t\!=\!\infty$, by putting
\bd
\Delta^2=Q\minus R^2\plus \eta^2 t/\alpha
\ed
with $Q$ and $R$ given by our previous exact result
(\ref{eq:QRDRTonline}),
one obtains
\be
E_{\rm g}=\frac{1}{\pi}\arccos\left[\frac{R}{\sqrt{Q}}\right]
~~~~~~~~~~~~~~~~~~
E_{\rm t}
=\frac{1}{2}-\frac{1}{2}\int\!Dy~
{\rm erf}\left[\frac{|y|R\plus\eta t/\alpha}{\Delta\sqrt{2}}\right]
\label{eq:errorsonline}
\ee

\begin{figure}[ht]
\centering
\vspace*{65mm}
\hbox to
\hsize{\hspace*{5mm}\includegraphics{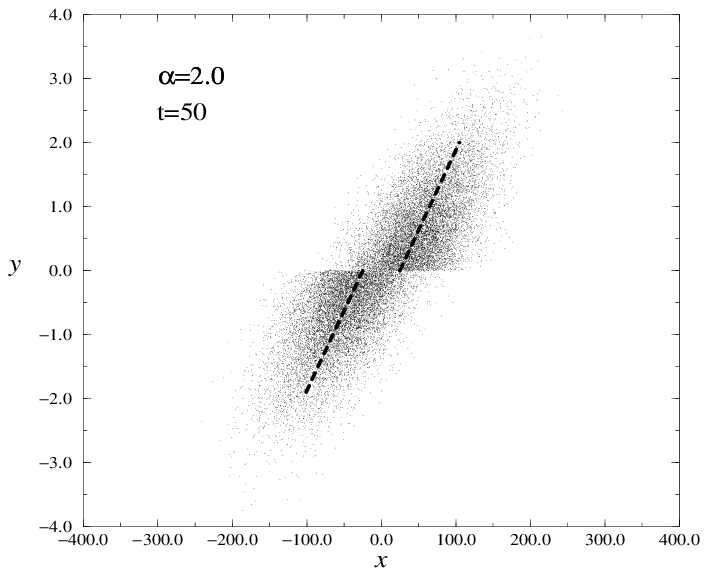}\hspace*{-5mm}}
\vspace*{55mm}
\hbox to
\hsize{\hspace*{5mm}\includegraphics{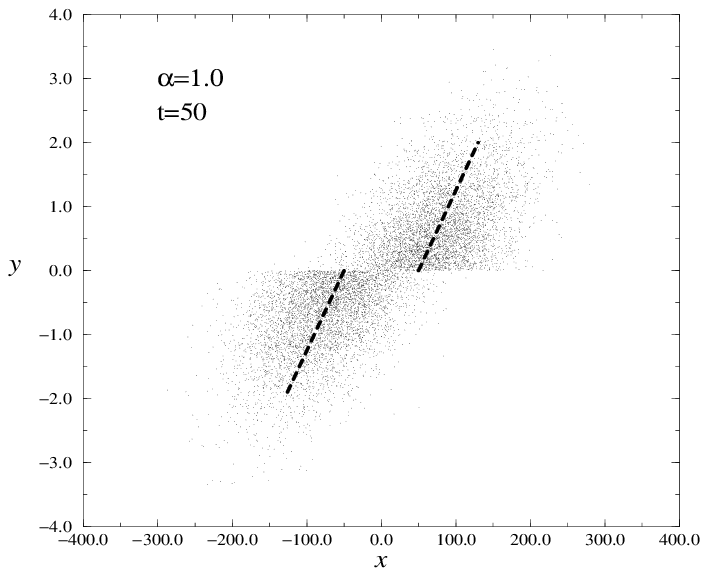}\hspace*{-5mm}}
\vspace*{55mm}
\hbox to
\hsize{\hspace*{5mm}\includegraphics{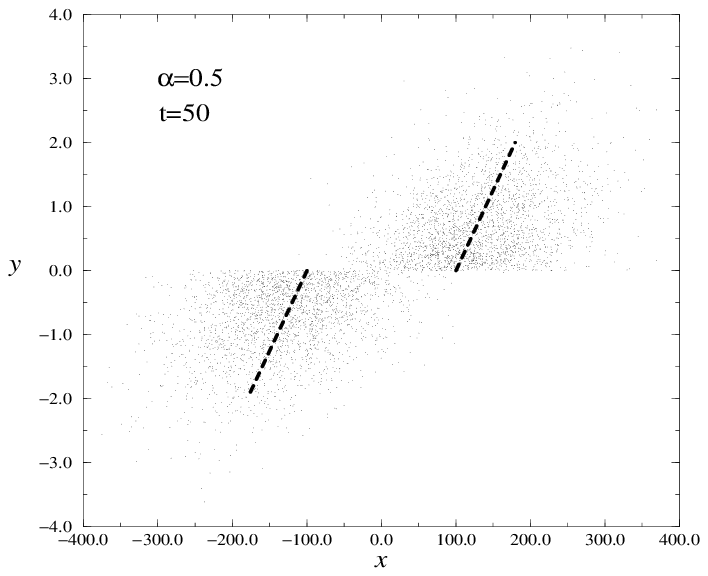}\hspace*{-5mm}}
\vspace*{-5mm}
\caption{Comparison between simulation results for on-line Hebbian
learning (system size $N=10,000$) and dynamical replica theory,
for $\eta=1$ and $\alpha\in\{0.5,1.0,2.0\}$. Dots: local fields
$(x,y)=(\bJ\inn\bxi,\bB\inn\bxi)$ (calculated for questions in the
training set),
at time  $t=50$. Dashed lines: conditional average of student field
$x$ as a function of $y$, as predicted by the theory,
$\overline{x}(y)=Ry+(\eta t/\alpha)\sgn(y)$.}
\label{fig:hebbPxy}
\end{figure}

\begin{figure}[ht]
\centering
\vspace*{55mm}
\hbox to
%compdisthebb_a4.eps
\hsize{\hspace*{8mm}\includegraphics{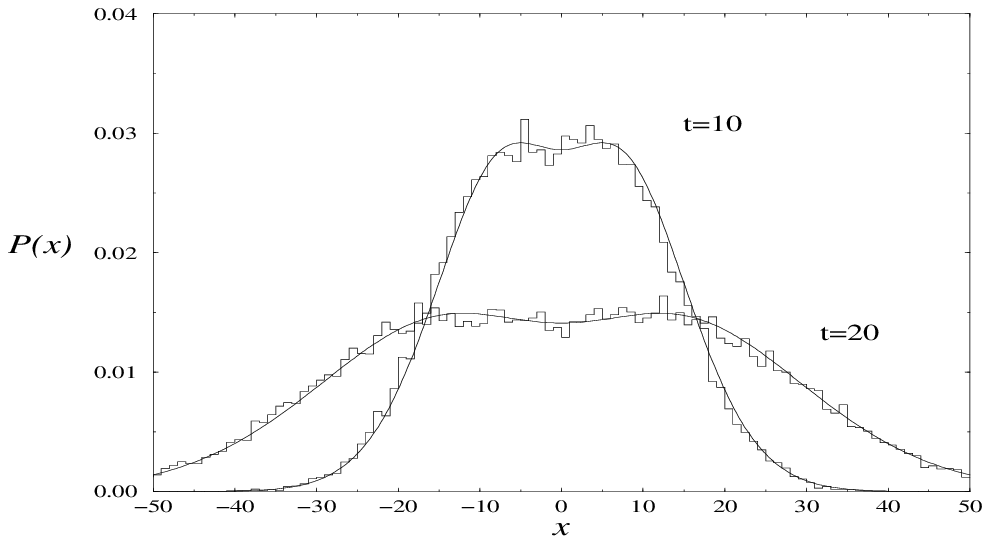}\hspace*{-8mm}}
\vspace*{52mm}
\hbox to
%compdisthebb_a1.eps
\hsize{\hspace*{8mm}\includegraphics{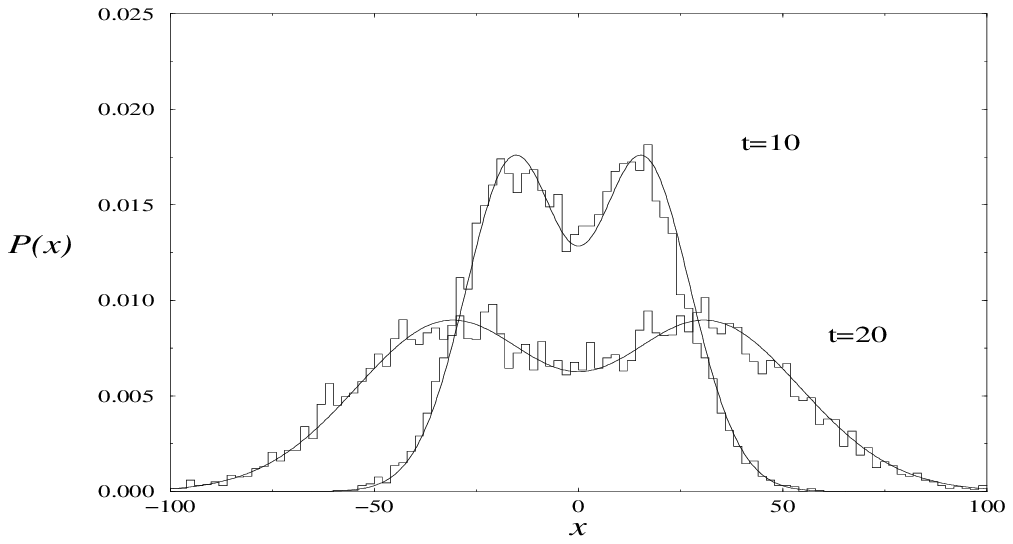}\hspace*{-8mm}}
\vspace*{52mm}
\hbox to
%compdisthebb_a025.eps
\hsize{\hspace*{8mm}\includegraphics{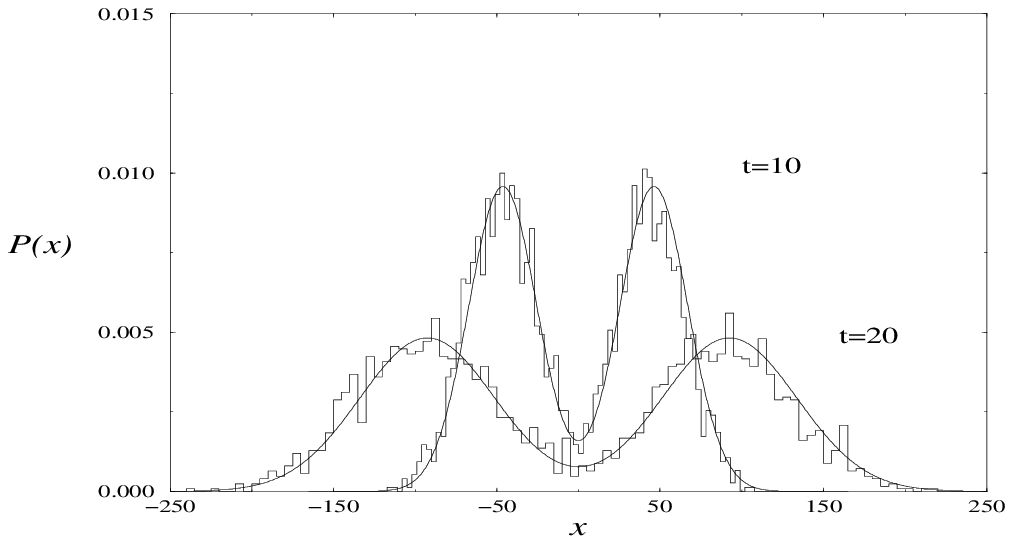}\hspace*{-8mm}}
\vspace*{-2mm}
\caption{Simulations of on-line Hebbian learning with $\eta=1$ and $N=10,000$.
Histograms: student field distributions measured at $t=10$ and $t=20$.
Lines: theoretical predictions for student field distributions.
$\alpha=4$ (upper), $\alpha=1$ (middle), $\alpha=0.25$
(lower).}
\label{fig:hebbPx}
\end{figure}

\noindent
We can also calculate
the student field distribution $P(x)=\int\!Dy ~P[x|y]$, giving
\bd
P(x)=\frac{e^{-\frac{1}{2}[x+\frac{\eta t}{\alpha}]^2/(\Delta^2\plus R^2)}}
{2\sqrt{2\pi (\Delta^2\!\plus R^2)}}
\left[1\minus
{\rm erf}\left(\frac{R[x\plus\frac{\eta t}{\alpha}]}{\Delta\sqrt{2(\Delta^2\!\plus R^2)}}
\right)\right]
~~~~~~~~~~~~~~~~~~~~~~~~~~~~~~~~~~~
\ed
\be
~~~~~~~~~~~~~~~~~~~~~~~~~~~~~~~~~~~
+\frac{e^{-\frac{1}{2}[x-\frac{\eta t}{\alpha}]^2/(\Delta^2\plus
R^2)}}{2\sqrt{2\pi (\Delta^2\!\plus R^2)}}
\left[1\plus {\rm erf}\left(\frac{R[x\minus \frac{\eta t}{\alpha}]}{\Delta\sqrt{2(\Delta^2\!\plus R^2)}}\right)\right]
\label{eq:Hebbstudentdist}
\ee
\vsp

In figure \ref{fig:hebberrors} we compare the
predictions
for the generalization and training errors ({\ref{eq:errorsonline})
of the approximate solution of our equations with the results
obtained from numerical simulations of on-line Hebbian learning
for $N=10,000$
(initial state: $Q_0=1$, $R_0=0$; learning rate: $\eta=1$).
All curves show
excellent agreement between
theory and
experiment.  For $E_{\rm g}$ this is
guaranteed by the exactness of our theory for $Q$ and $R$; the
agreement found for $E_{\rm t}$ is more surprising, in that these
predictions are obtained from a simple approximation of the
solution of our equations.
We also compare the theoretical predictions made for the
distribution $P[x|y]$ with the results of numerical simulations.
This
is done in figure \ref{fig:hebbPxy}, where we show the fields as
observed at time $t=50$ in simulations ($N=10,000$, $\eta=1$, $R_0=0$,
$Q_0=1$) of on-line Hebbian learning, for three different values of
$\alpha$. In the same figure  we draw (as dashed lines) the
theoretical prediction (\ref{eq:Hebbxaverage})
for the $y$-dependent average of the
conditional $x$-distribution $P[x|y]$.
Finally we compare the student field distribution $P(x)$, as
observed in simulations
of on-line Hebbian learning ($N=10,000$, $\eta=1$, $R_0=0$,
$Q_0=1$)
with our prediction (\ref{eq:Hebbstudentdist}). The result is
shown in figure \ref{fig:hebbPx}, for $\alpha\in\{4,1,0.25\}$.
In all cases the agreement between theory and experiment, even for the
approximate solution of our equations, is quite satisfactory.

\clearpage
\section{General Approximation Schemes}

All three approximation schemes presented in this section aim at
providing alternatives to calculating the effective measure $M[x|y]$
at each time step from the functional saddle-point equation.
Since this calculation cannot (yet) be done analytically, it constitutes a
significant numerical obstacle in working
out the predictions of our theory.
Each scheme preserves both
 normalisation and symmetries of the probability density $P[x,y]$ and its
marginals, as well as the relation $\int\!dx~P[x|y]\Phi[x,y]=0$
for all $y$.
In the first two approximation schemes, a large $\alpha$ expansion
and a 
conditionally-Gaussian saddle-point approximation, all Gaussian
integrals 
representing the disorder in
the problem can be done analytically; this leads to a
significant reduction in CPU time when solving our equations numerically
(especially the large $\alpha$ approximation is
extremely simple and fast, as it does not even involve
a saddle-point equation for $q$).
We only work out the equations for on-line learning; the  batch laws
follows as usual upon expanding the equations 
in powers of $\eta$ and retaining only the linear terms.

\subsection{Large $\alpha$ Approximation}

Our first approximation scheme is obtained upon taking into
account the finite nature of the training set (i.e. the disordered
nature of the dynamics) in first non-trivial
order. The amount of disorder is effectively measured by the
parameter $B$, or, equivalently, by the deviation of the value of
the spin-glass order parameter $q$ from its naive value $R^2/Q$.
Putting $B=0$ in the saddle-point equation (\ref{eq:summarysaddle_M})
immediately gives $\lim_{B\to 0}M[x|y]=P[x|y]$, so we
write
\be
M[x|y]=P[x|y]~[1+\sum_{\ell>0}B^\ell m_\ell[x|y]],~~~~~~~~~~
\int\!dx~P[x|y] m_\ell[x|y]=0
\label{eq:approx1_ansatz}
\ee
Upon inserting (\ref{eq:approx1_ansatz}) as an {\em ansatz}
into the saddle-point equation (\ref{eq:summarysaddle_M})
, one easily shows that
\be
M[x|y]=P[x|y]~e^{-\frac{1}{2}B^2[x- \overline{x}(y)]^2
+\frac{1}{2}B^2\left[\overline{x^2}(y)-\overline{x}(y)^2\right]+\order(B^3)}
\label{eq:approx1_M}
\ee
with the abbreviations
\bd
\overline{x}(y)=\int\!dx~P[x|y]x
~~~~~~~~~~~~~~
\overline{x^2}(y)=\int\!dx~P[x|y]x^{2}
\ed
(the second $\order(B^2)$ term in the exponent of (\ref{eq:approx1_M}), being
independent of $x$, just reflects the normalisation requirements).
This result enables us, in turn, to expand the function $\Phi[x,y]$
which controls the non-trivial term in our diffusion equation for
$P[x|y]$.
Note that from the definition of $B$ it follows that
$Q(1\minus q)=\frac{1}{2}B^{-2}[\sqrt{1\plus 4B^2(Q\minus R^2)}\minus
1]$,
which gives
\bd
\Phi[x,y]=
\frac{x-\overline{x}(y)}{Q\minus R^2}
+\order(B^2)
\ed
With this expression we can write our approximate equations in explicitly
closed form (i.e. without any remaining saddle-point equations).
The relevant scalar functions become
\be
U=\frac{\bra \G[x,y][x\minus \overline{x}(y)]\ket}{Q\minus
R^2}~~~~~~~~
V=\bra x \G[x,y]\ket~~~~~~~~
W=\bra y \G[x,y]\ket~~~~~~~~
Z=\bra \G^2[x,y]\ket
\label{eq:approx1_fouraverages}
\ee
For on-line learning we find:
\be
\frac{d}{dt}Q=2\eta V +\eta^2 Z
~~~~~~~~~~~~~~~~
\frac{d}{dt}R =
\eta W
\label{eq:approx1_onlinedQdtdRdt}
\ee
\bd
\frac{d}{dt}P[x|y]=\frac{1}{\alpha}\int\!dx^\prime P[x^\prime|y]\left[
\delta[x\minus
x^\prime\minus \eta\G[x^\prime\!,y]]\minus \delta[x\minus
x^\prime]\right]
- \eta\frac{\partial}{\partial x} \left\{\room
P[x|y]\left[ U(x\minus Ry) \plus Wy\right]\right\}
\ed
\be
+ \frac{1}{2}\eta^2 Z
\frac{\partial^2}{\partial x^2}P[x|y]
- \eta\left[\frac{V\minus RW}{Q\minus R^2}-U \right]
\frac{\partial}{\partial x}\left\{\room P[x|y]
[x\minus \overline{x}(y)]
\right\}
\label{eq:approx1_onlinedPdt}
\ee
From the solution of the above equations follow, as always, the training- and
generalization
errors  $E_{\rm t}=\int\!Dy dx~P[x|y]\theta[-xy]$
and $E_{\rm g}=\pi^{-1}\arccos[R/\sqrt{Q}]$.
The resulting theory is obviously exact in the limit $\alpha\to\infty$
(see \cite{CoolenSaad1}), by construction.

\subsection{Conditionally-Gaussian Approximation}

Our basic idea here is a variational approach to solving the functional
saddle-point problem (valid for any $\alpha$), i.e.
to carry out the functional extremisation only
within the restricted family of conditionally Gaussian measures
$M[x|y]$ (which, together with $q$, characterises the saddle-point):
\bd
M[x|y]=
\frac{e^{-\frac{1}{2}[x-\overline{x}(y)]^2/\sigma^2(y)}}{\sigma(y)\sqrt{2\pi}}
\ed
Note that this does not imply the stronger statement that $P[x|y]$
itself is taken 
to be of a conditionally-Gaussian form (as in the case of the approximation used for
on-line Hebbian learning).
Extremisation of the original replica-symmetric functional $\Psi[q,\{M\}]$
(see \cite{CoolenSaad1}) within the conditionally-Gaussian family of functions
 results in the requirement  that
the two $y$-dependent moments $\overline{x}(y)$ and $\sigma^2(y)$
be given by
\bd
\overline{x}(y)=\int\!dx~x P[x|y],~~~~~~~~~~
\Delta^2(y)=\int\!dx~x^2 P[x|y]-\overline{x}^2(y)
=\sigma^2(y)\plus B^2\sigma^4(y)
\ed
Now we can again calculate all relevant averages which involve the effective measure
$M[x|y]$ exactly. In particular:
\bd
\bra x\ket_\star = \overline{x}(y)+zB\sigma^2(y)
~~~~~~~~~~~~~~~~
B=\frac{\sqrt{qQ\minus R^2}}{Q(1\minus q)}
~~~~~~~~~~~~~~~~
\Phi[x,y]=
\frac{e^{-\frac{1}{2}[x-\overline{x}(y)]^2/\Delta^2(y)}}{\Delta(y)\sqrt{2\pi}
P[x|y]}
\frac{(x-\overline{x}(y))\sigma^2(y)}{Q(1\minus q)\Delta^2(y)}
\ed
For on-line learning this results in the following approximated theory:
\bd
U=\int\!DyDu\left\{
\frac{u\sigma^2(y)\G[\overline{x}(y)\plus u\Delta(y),y]}{Q(1\minus q)\Delta(y)}
\right\}
\ed
\be
\room
V=\bra x \G[x,y]\ket~~~~~~~~~~
W=\bra y \G[x,y]\ket~~~~~~~~~~
Z=\bra \G^2[x,y]\ket
\label{eq:approx2_fouraverages}
\ee
\be
\room
\frac{d}{dt}Q=2\eta V +\eta^2 Z
~~~~~~~~~~~~~~~~
\frac{d}{dt}R =
\eta W
\label{eq:approx2_onlinedQdtdRdt}
\ee
\bd
\room
\frac{d}{dt}P[x|y]=\frac{1}{\alpha}\int\!dx^\prime P[x^\prime|y]\left[
\delta[x\minus
x^\prime\minus \eta\G[x^\prime\!,y]]\minus \delta[x\minus
x^\prime]\right]
- \eta\frac{\partial}{\partial x} \left\{\room
P[x|y]\left[ U(x\minus Ry) \plus Wy\right]\right\}
~~~~~~~~~~
\ed
\be
\room
+ \frac{1}{2}\eta^2 Z
\frac{\partial^2}{\partial x^2}P[x|y]
- \frac{\eta\sigma^2(y)\left[V\! \minus RW\! \minus (Q\minus R^2)U \right]}
{\sqrt{2\pi}Q(1\minus q)\Delta^5(y)}
\left[\Delta^2(y)\minus (x\minus \overline{x}(y))^2\right]
e^{-\frac{1}{2}[x-\overline{x}(y)]^2/\Delta^2(y)}
\label{eq:approx2_onlinedPdt}
\ee
The remaining order parameter $q$ is calculated at each time-step by
solving
\bd
\bra (x\minus Ry)^2\ket
+(qQ\minus R^2)(1\minus \frac{1}{\alpha})
=\left[2\frac{qQ\minus R^2}{Q(1\minus q)}\plus 1\right]
\int\!Dy~ \sigma^2(y)
\ed
From the solution of these equations follow the training- and
generalization
errors  $E_{\rm t}=\int\!Dy dx~P[x|y] \theta[-xy]$
and $E_{\rm g}=\pi^{-1}\arccos[R/\sqrt{Q}]$.

\subsection{Partially Annealed Approximation}

In order to construct our third and final approximation we return
to an earlier stage of the derivation of the present formalism
(see \cite{CoolenSaad1}), and
rewrite the functional saddle-point equation in a form where the
replica limit $n\to 0$ has not yet been taken, i.e.
\bd
{\rm for~all}~x,y:~~~~~~~~~~~~~~
P[x|y]=\frac{\int\!Dz~M_n[x|y]e^{Bz[x-\overline{x}(y)]}
\left[\int\!dx^\prime~M_n[x^\prime|y]e^{Bz[x^\prime -\overline{x}(y)]}\right]^{n-1}}
{\int\!Dz \left[\int\!dx^\prime~M_n[x^\prime|y]e^{Bz[x^\prime -\overline{x}(y)]}\right]^{n}}
\ed
with $\overline{x}(y)=\int\!dx~xP[x|y]$.
In our full (quenched disorder) calculation we find ourselves with the effective measure
 $M[x|y]=\lim_{n\to
0}M_n[x|y]$.
In contrast, an alternative calculation, whereby the quenched average over all
training sets would have been replaced by an annealed average over
all training sets, would have led us to the value $n=1$ rather than $n=0$:
$M[x|y]=M_1[x|y]$. We can now define in a natural way an annealed
approximation of our theory upon replacing the complicated $n=0$ functional
saddle-point equation (\ref{eq:summarysaddle_M}) by the much simpler $n=1$
version:
\bd
P[x|y]=\frac{\int\!Dz~M[x|y]e^{Bz[x-\overline{x}(y)]}}
{\int\!Dz \int\!dx^\prime~M[x^\prime|y]e^{Bz[x^\prime-\overline{x}(y)]}}
\ed
The $z$-integrations can immediately be carried out, and the
resulting equation solved for $M[x|y]$, giving:
\be
M[x|y]=\frac{P[x|y]~e^{-\frac{1}{2}B^2[x-\overline{x}(y)]^2}}
{\int\!dx^\prime
~P[x^\prime|y]~e^{-\frac{1}{2}B^2[x^\prime-\overline{x}(y)]^2}},
\label{eq:approx3_M}
\ee
Averages involving the effective measure $M[x|y]$ are thus written
explicitly in terms of $P[x|y]$, and
we are left with the following approximate theory:
\be
\room
U=\bra \Phi[x,y]\G[x,y]\ket~~~~~~~~~
V=\bra x \G[x,y]\ket~~~~~~~~~
W=\bra y \G[x,y]\ket~~~~~~~~~
Z=\bra \G^2[x,y]\ket
\label{eq:approx3_fouraverages}
\ee
\be
\room
\frac{d}{dt}Q=2\eta V +\eta^2 Z
~~~~~~~~~~~~~~~~
\frac{d}{dt}R =
\eta W
\label{eq:approx3_onlinedQdtdRdt}
\ee
\bd
\room
\frac{d}{dt}P[x|y]=\frac{1}{\alpha}\int\!dx^\prime P[x^\prime|y]\left[
\delta[x\minus
x^\prime\minus \eta\G[x^\prime\!,y]]\minus \delta[x\minus
x^\prime]\right]
- \eta\frac{\partial}{\partial x} \left\{\room
P[x|y]\left[ U(x\minus Ry) \plus Wy\right]\right\}
\ed
\be
\room
+ \frac{1}{2}\eta^2 Z
\frac{\partial^2}{\partial x^2}P[x|y]
- \eta\left[V\! \minus RW\! \minus (Q\minus R^2)U \right]
\frac{\partial}{\partial x}\left\{\room
P[x|y]\Phi[x,y]
\right\}
\label{eq:approx3_onlinedPdt}
\ee
with
\bd
\Phi[X,y]=\frac{1}{Q(1\minus q)}\int\!Dz\left\{
 \frac{\int\!dx~P[x|y]~e^{-\frac{1}{2}[B(x-\overline{x}(y))-z]^2-\frac{1}{2}[B(X-\overline{x}(y))-z]^2}(X-x)}
{\left[\int\!dx~P[x|y]~e^{-\frac{1}{2}[B(x-\overline{x}(y))-z]^2}\right]^2}
\right\}
\ed
As always, $B=\sqrt{qQ\minus R^2}/Q(1\minus q)$.
The remaining spin-glass order parameter $q$ is calculated at each time-step by solving
\bd
\bra (x\minus Ry)^2\ket
+(qQ\minus R^2)(1\minus \frac{1}{\alpha})
=\left[2(qQ\minus R^2)^\frac{1}{2}\plus \frac{1}{B}\right]
\int\!DyDz~ z\left\{
 \frac{\int\!dx~P[x|y]~e^{-\frac{1}{2}[B(x-\overline{x}(y))-z]^2}x}
{\int\!dx~P[x|y]~e^{-\frac{1}{2}[B(x-\overline{x}(y))-z]^2}}
\right\}
\ed
From the solution of the above equations follow the training- and generalization
errors  $E_{\rm t}=\bra \theta[-xy]\ket$
and $E_{\rm g}=\pi^{-1}\arccos[R/\sqrt{Q}]$.
It should be emphasised that the present approximation is not
equivalent to 
(and should be more accurate than) a full
annealed treatment of the disorder in the problem; the latter would
have affected not only the equation for $M[x|y]$ but also the
saddle-point equation for $q$ (hence the name {\em partially} 
annealed approximation).

\section{Non-Hebbian Rules: Theory versus Simulations}

\begin{figure}[t]
\centering
\vspace*{60mm}
\hbox to\hsize{\hspace*{18mm}\includegraphics{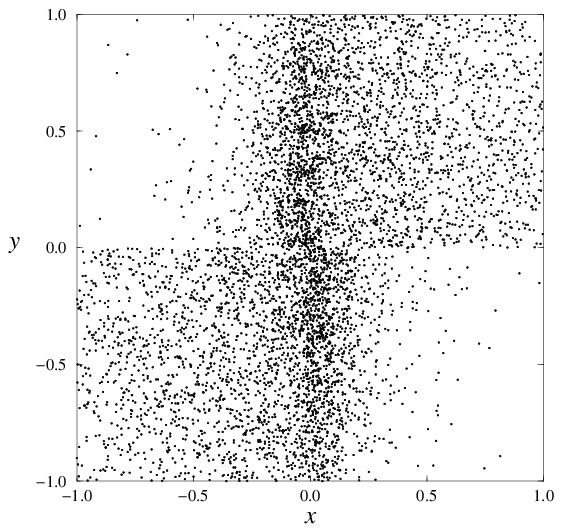}\hspace*{-18mm}}
\vspace*{-4.8mm}
\hbox to\hsize{\hspace*{78mm}\includegraphics{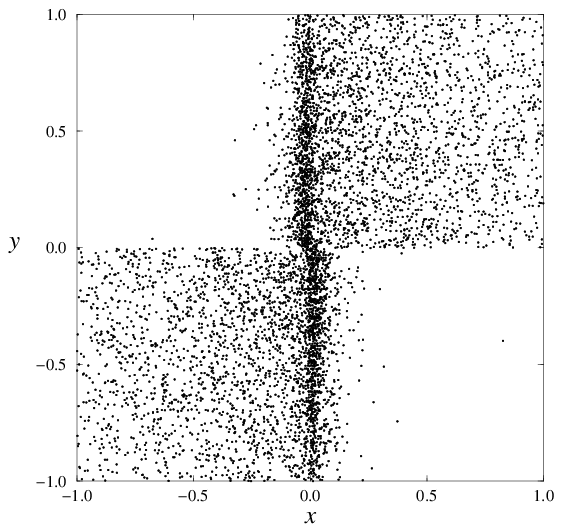}\hspace*{-78mm}}
\vspace*{52mm}
\hbox to\hsize{\hspace*{18mm}\includegraphics{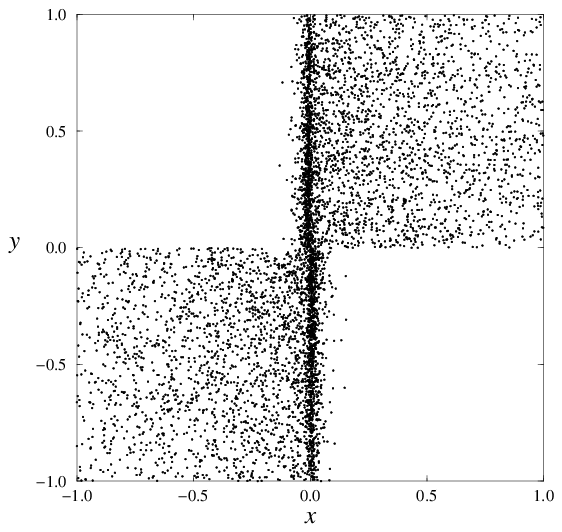}\hspace*{-18mm}}
\vspace*{-4.8mm}
\hbox to\hsize{\hspace*{78mm}\includegraphics{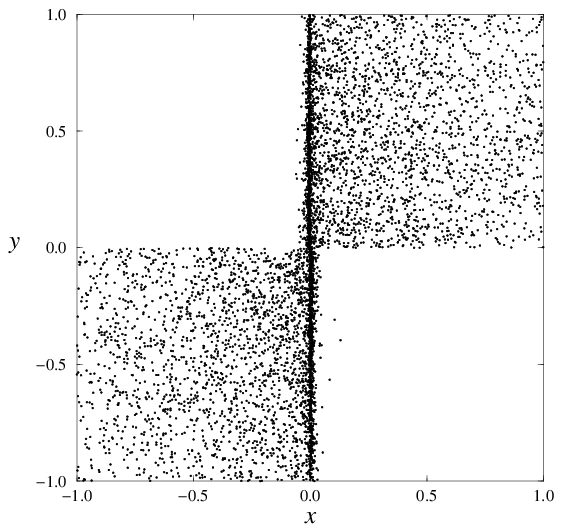}\hspace*{-78mm}}
\vspace*{-5mm}
\caption{Numerical simulations of on-line Adatron learning, with
$N\!=\!10,\!000$, $\alpha\!=\!1$ and $\eta\!=\!\frac{1}{2}$.
The scatter plots show the observed student and teacher fields
$(x,y)\!=\!(\bJ\inn\bxi,\bB\inn\bxi)$ at times $t\!=\!5$ (upper left),
$t\!=\!10$ (upper right), $t\!=\!15$ (lower left) and $t\!=\!20$ (lower right), as measured during simulations
for the data in the training set $\set$, drawn as points in the $(x,y)$ plane. Note the development over time of
an increasingly narrow `ridge' along the line $x\!=\!0$.}
\label{fig:adatdelta}
\end{figure}

Henceforth we will always assume  initial states with specified
values for $R_0$ and $Q_0$ but without correlations with the training
set, i.e.
\bd
P_0[x|y]=\frac{e^{-\frac{1}{2}[x-R_0 y]^2/(Q_0-R_0^2)}}{\sqrt{2\pi(Q_0-R_0^2)}}
\ed
This implies that the student could initially have some knowledge
of the rule to be learned, if we wish, but will never know
beforehand about the composition of the training  set.
We will inspect the learning dynamics generated upon using two of the
most common non-Hebbian
(error-correcting) learning
rules:
\be
\begin{array}{lll}
{\rm Perceptron:}  && \G[x,y]=\sgn(y)\theta[-xy]\\[3mm]
{\rm AdaTron:}     && \G[x,y]=|x|\sgn(y)\theta[-xy]
\end{array}
\label{eq:nonhebbian_rules}
\ee
Note that in the case of AdaTron learning  the cases $\eta\leq 1$ and
$\eta>1$ give rise to qualitatively different behaviour of the
first term in the diffusion equation (\ref{eq:summaryonlinedPdt}). 
For $\eta<1$ the learning process, aiming at the 
situation where $xy>0$ never occurs, remedies inappropriate
student fields by slowly moving them towards (but not immediately across)
the decision
boundary. For $\eta>1$ the adjustments made to the student fields
could move them well into the region at the other side of the decision boundary.
The case $\eta=1$ is special, in that changes to the student
fields tend to move them precisely onto the decision
boundary. The student field distribution consequently
develops  a
$\delta$-peak at the origin, in perfect agreement with what can be observed in
numerical simulations
(see  e.g. the figures referring to on-line AdaTron learning with $\eta=1$ in \cite{CoolenSaad1}):
\bd
\eta=1:~~~~~~~~~~
\frac{d}{dt}P[x|y]=
\frac{1}{\alpha}\left\{\room
\delta(x)\int\!dx^\prime~\theta[\minus x^\prime y] P[x^\prime|y]
-P[x|y]\theta[\minus xy]\right\}
+\ldots
\ed
In fact the same occurs for all $\eta\leq 1$: about
half of the probability weight of $P[x|y]$
will in due course become concentrated in an increasingly thin ridge along the decision
boundary $x=0$. This is illustrated in figure \ref{fig:adatdelta},
for $\eta=\frac{1}{2}$.
Since such a singular behaviour (although in principle
accurately described by our equations)
will be difficult to reproduce when solving the equations numerically, 
using  finite spatial  resolution, we will in this paper only deal
with the case of $\eta> 1$
for AdaTron learning.

\subsection{Large $\alpha$ and Conditionally-Gaussian Approximations}

\begin{figure}[t]
\centering
\vspace*{68mm}
\hbox to
\hsize{\hspace*{0mm}\includegraphics{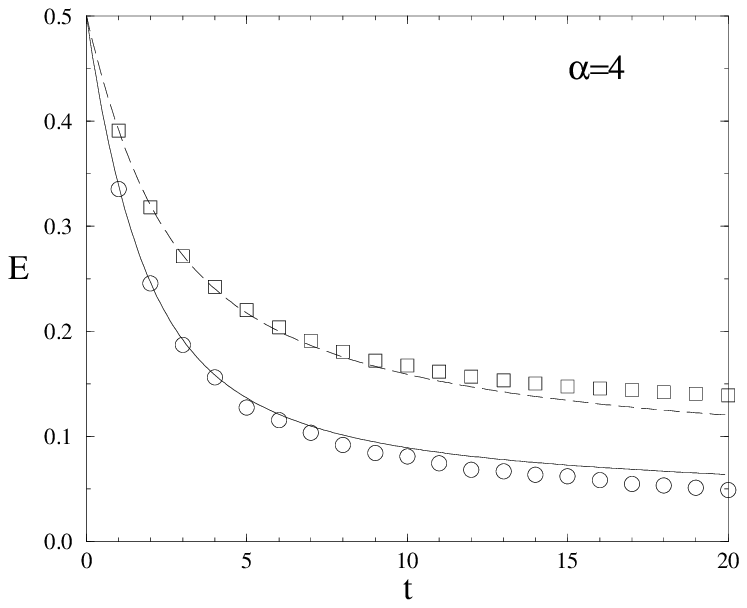}\hspace*{0mm}}
\vspace*{-5mm}
\hbox to
\hsize{\hspace*{80mm}\includegraphics{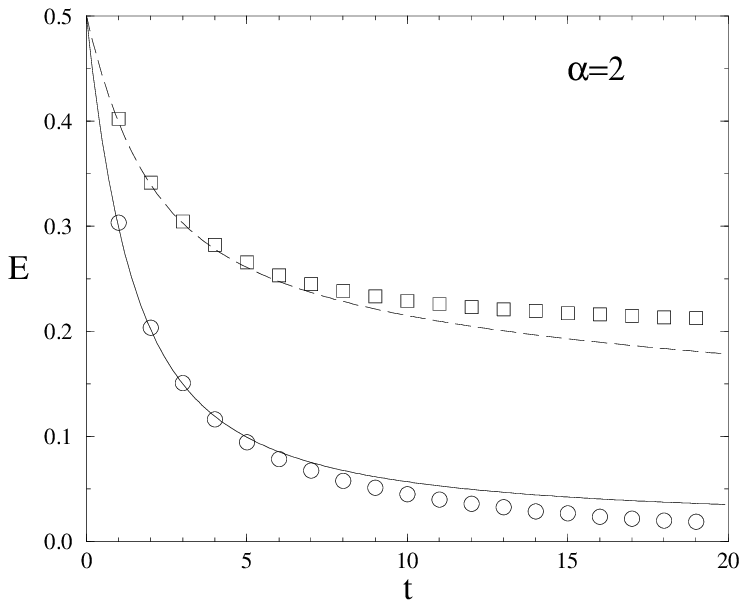}\hspace*{-80mm}}
\vspace*{60mm}
\hbox to
\hsize{\hspace*{0mm}\includegraphics{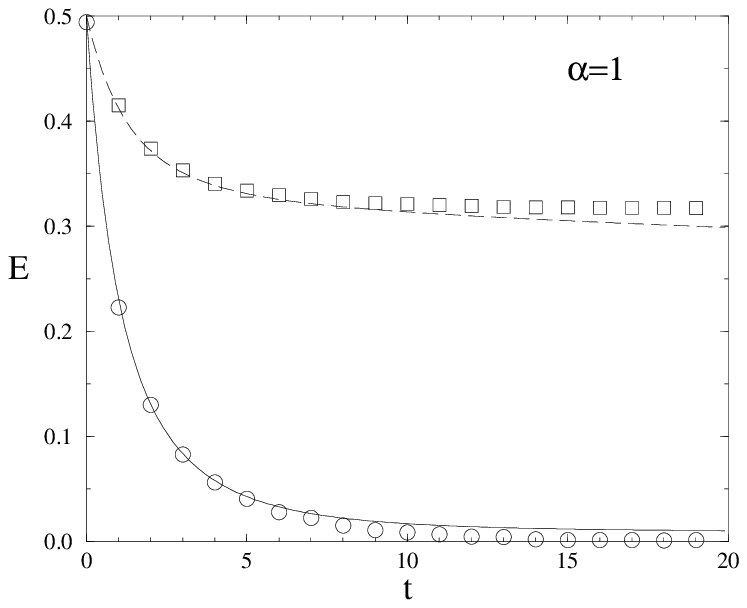}\hspace*{0mm}}
\vspace*{-5mm}
\hbox to
\hsize{\hspace*{80mm}\includegraphics{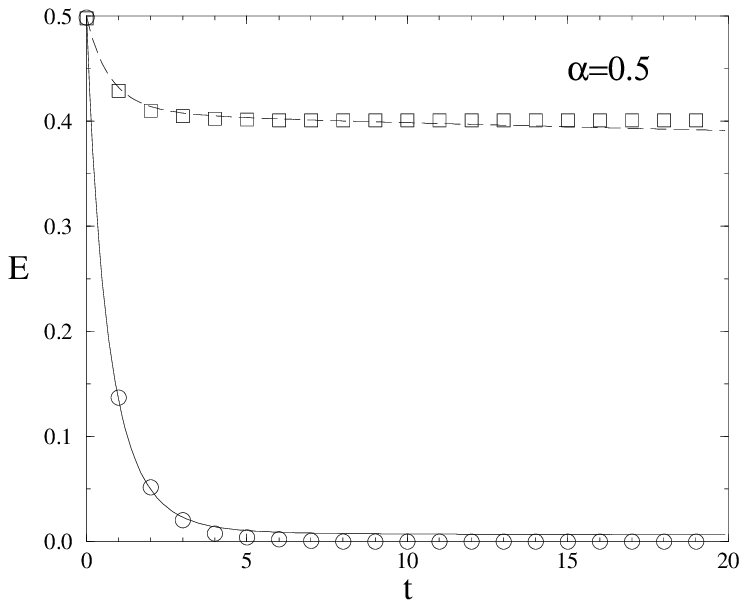}\hspace*{-80mm}}
\vspace*{-5mm}
\caption{Comparison between the large $\alpha$ approximation of the
theory and numerical simulations of on-line perceptron learning with
$N=10,000$ and $\eta=1$.
Markers: training errors $E_{\rm t}$ (circles) and generalisation errors
$E_{\rm g}$
(squares); 
finite size effects in the simulation data are of the
order of the marker size. 
Lines: theoretical predictions for training errors (solid) and
generalisation errors (dashed) as functions of time, according to the
approximated theory. Training set sizes:
$\alpha=4$ (upper left), $\alpha=2$ (upper right), $\alpha=1$ (lower
left), and $\alpha=0.5$ (lower right). }
\label{fig:approx1a}
\end{figure}

\begin{figure}[t]
\centering
\vspace*{68mm}
\hbox to
\hsize{\hspace*{0mm}\includegraphics{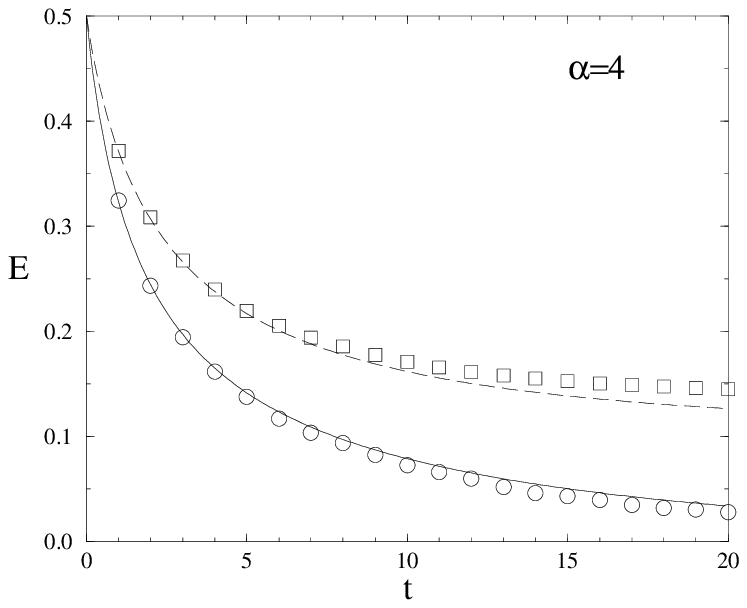}\hspace*{0mm}}
\vspace*{-5mm}
\hbox to
\hsize{\hspace*{80mm}\includegraphics{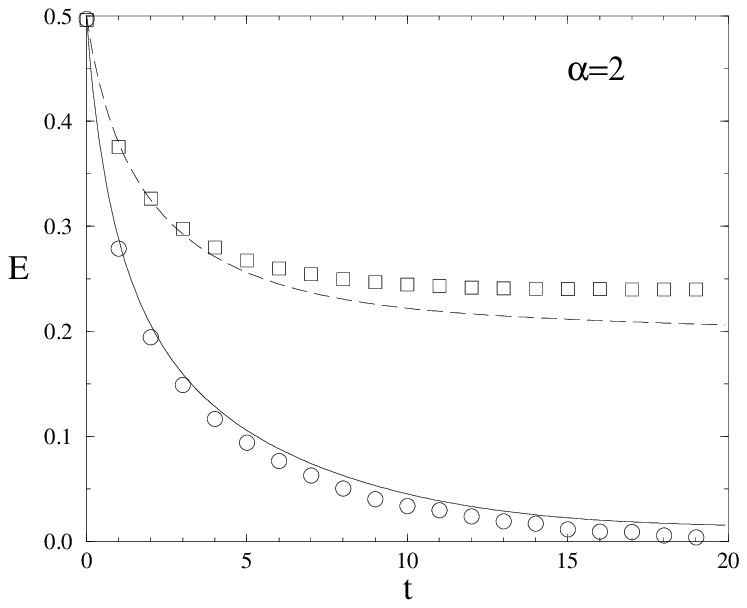}\hspace*{-80mm}}
\vspace*{60mm}
\hbox to
\hsize{\hspace*{0mm}\includegraphics{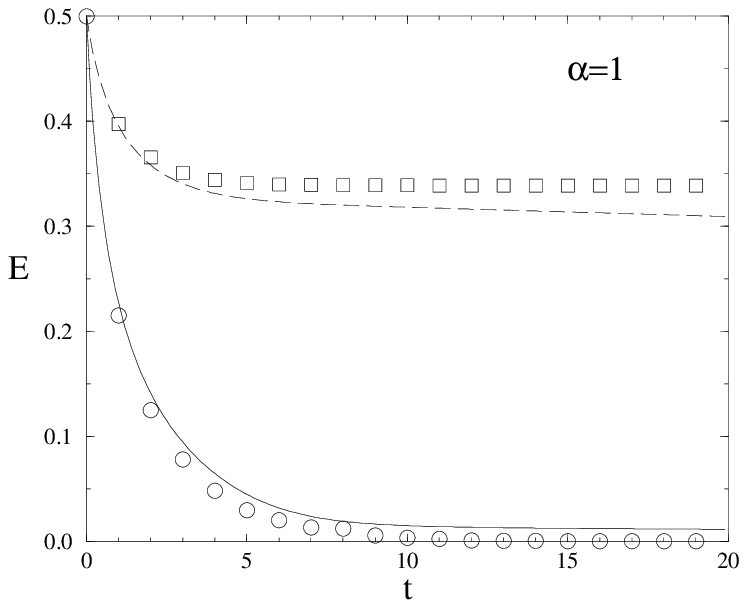}\hspace*{0mm}}
\vspace*{-5mm}
\hbox to
\hsize{\hspace*{80mm}\includegraphics{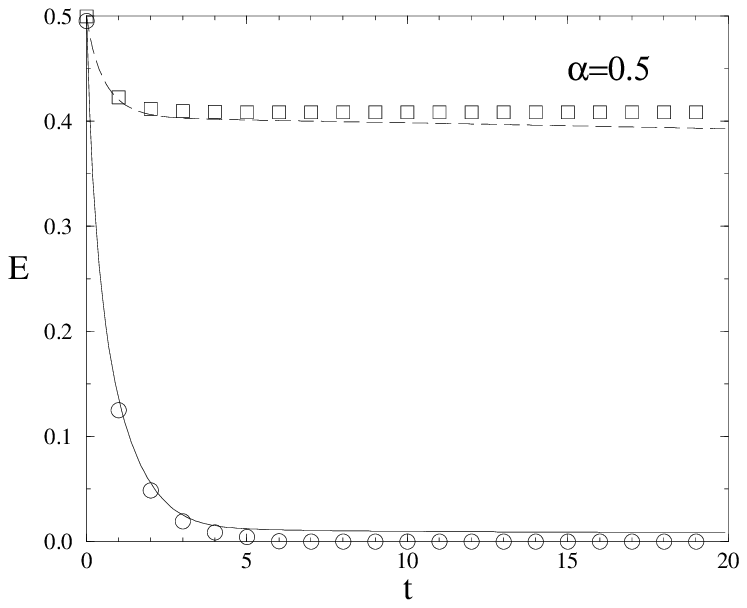}\hspace*{-80mm}}
\vspace*{-5mm}
\caption{Comparison between the large $\alpha$ approximation of the
theory and numerical simulations of on-line Adatron learning with
$N=10,000$ and $\eta=2$.
Markers: training errors $E_{\rm t}$ (circles) and generalisation errors
$E_{\rm g}$
(squares); finite size effects in the simulation data are of the
order of the marker size. 
Lines: theoretical predictions for training errors (solid) and
generalisation errors (dashed) as functions of time, according to the
approximated theory. Training set sizes:
$\alpha=4$ (upper left), $\alpha=2$ (upper right), $\alpha=1$ (lower
left), and $\alpha=0.5$ (lower right). }
\label{fig:approx1b}
\end{figure}

Our first approximated theory (the large $\alpha$ approximation) is very simple, with neither saddle-point
equations to be solved nor nested integrations. As a result, 
numerical solution of the macroscopic equations is straightforward
and fast.
In figures \ref{fig:approx1a} (on-line perceptron learning) and \ref{fig:approx1b}
(on-line Adatron learning) we compare
the results of solving the coupled equations
(\ref{eq:approx1_fouraverages},\ref{eq:approx1_onlinedQdtdRdt},\ref{eq:approx1_onlinedPdt})
numerically for finite values of $\alpha$, plotting the generalisation- and training errors as
functions of time, with results obtained from performing numerical
simulations. 
 As could have been expected, the large $\alpha$ approximation under-estimates
the amount of disorder in the learning process, which immediately
translates into under-estimation of the gap between $E_{\rm t}$
and $E_{\rm g}$ (which is its fingerprint).
It is also clear from these figures that, although at any given
time the quality of the predictions of this approximation does improve when $\alpha$ 
increases (as indeed it should), and although there is surely qualitative agreement,
reliably accurate 
quantitative statements on the values of the training- and
generalisation errors are confined to the
regime $\eta t\leq \alpha$. Yet, surprisingly, the
agreement obtained is very good, even for $\eta t>\alpha$. Apparently the present
approximation does still capture the main characteristics of the
(non-Gaussian) joint field distribution. This is illustrated
quite clearly and explicitly in figures \ref{fig:approx1c}
and \ref{fig:approx1d}, where we compare for a fixed time $t=10$
the student and teacher
fields as measured during
numerical simulations (for $N=10,000$, drawn as dots in the
$(x,y)$ plane) for the $p=\alpha N$ questions $\bxi^\mu$ in the training
set $\set$, to the theoretical predictions for the joint field distribution
$P[x,y]$ (drawn as contour
plots).
We will not at this stage attempt to explain the surprising effectiveness 
of the large $\alpha$ approximation for small values of $\alpha$ (note
that figures \ref{fig:approx1a} and \ref{fig:approx1b} even suggest an 
increase in accurateness as $\alpha$ is lowered below $\alpha=1$).  
This would require a systematic mathematical analysis of the non-linear diffusion
equation (\ref{eq:approx1_onlinedPdt}), which we consider to be beyond
the scope of the present paper.  
\vsp

The conditionally-Gaussian  approximation again involves no
nested integrals, and its equations can therefore still be solved numerically
in a reasonably fast way,
but it does already require the solution (at each infinitesimal time step)
of a scalar
saddle-point equation to determine the spin-glass order parameter $q$.
Approximations of this type  work
extremely well for the simple Hebbian learning 
\clearpage

\begin{figure}[ht]
\centering
\vspace*{225mm} \hbox to
\hsize{\hspace*{-21mm}\includegraphics{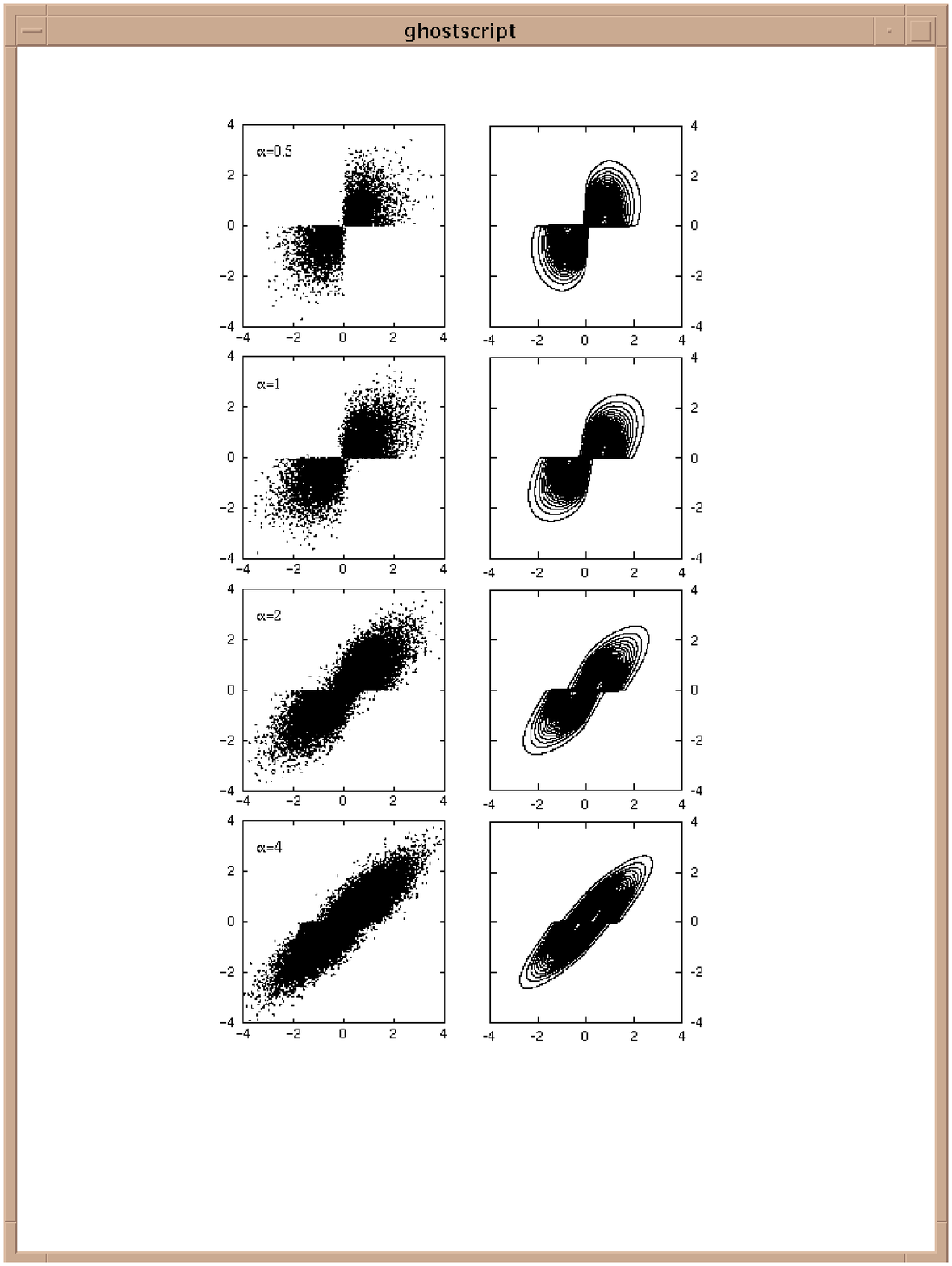}\hspace*{21mm}}
\vspace*{-45mm}
\caption{Comparison between the large $\alpha$ approximation of the
theory and numerical simulations of on-line Perceptron learning, with
$N\!=\!10,000$ and $\eta\!=\!1$.
Scatter plots (left): observed student and teacher fields
$(x,y)\!=\!(\bJ\inn\bxi,\bB\inn\bxi)$ as measured at time
$t\!=\!10$ during simulations,
for the data in $\set$, drawn in the $(x,y)$ plane. Contour plots  (right): corresponding
predictions for the joint field distribution $P[x,y]$, according to the
approximated theory. Training set sizes:
$\alpha\!=\!0.5,1,2,4$ (from top to bottom).}
\vspace*{-5mm}
\label{fig:approx1c}
\end{figure}

\clearpage

\begin{figure}[ht]
\centering
\vspace*{225mm} \hbox to
\hsize{\hspace*{-21mm}\includegraphics{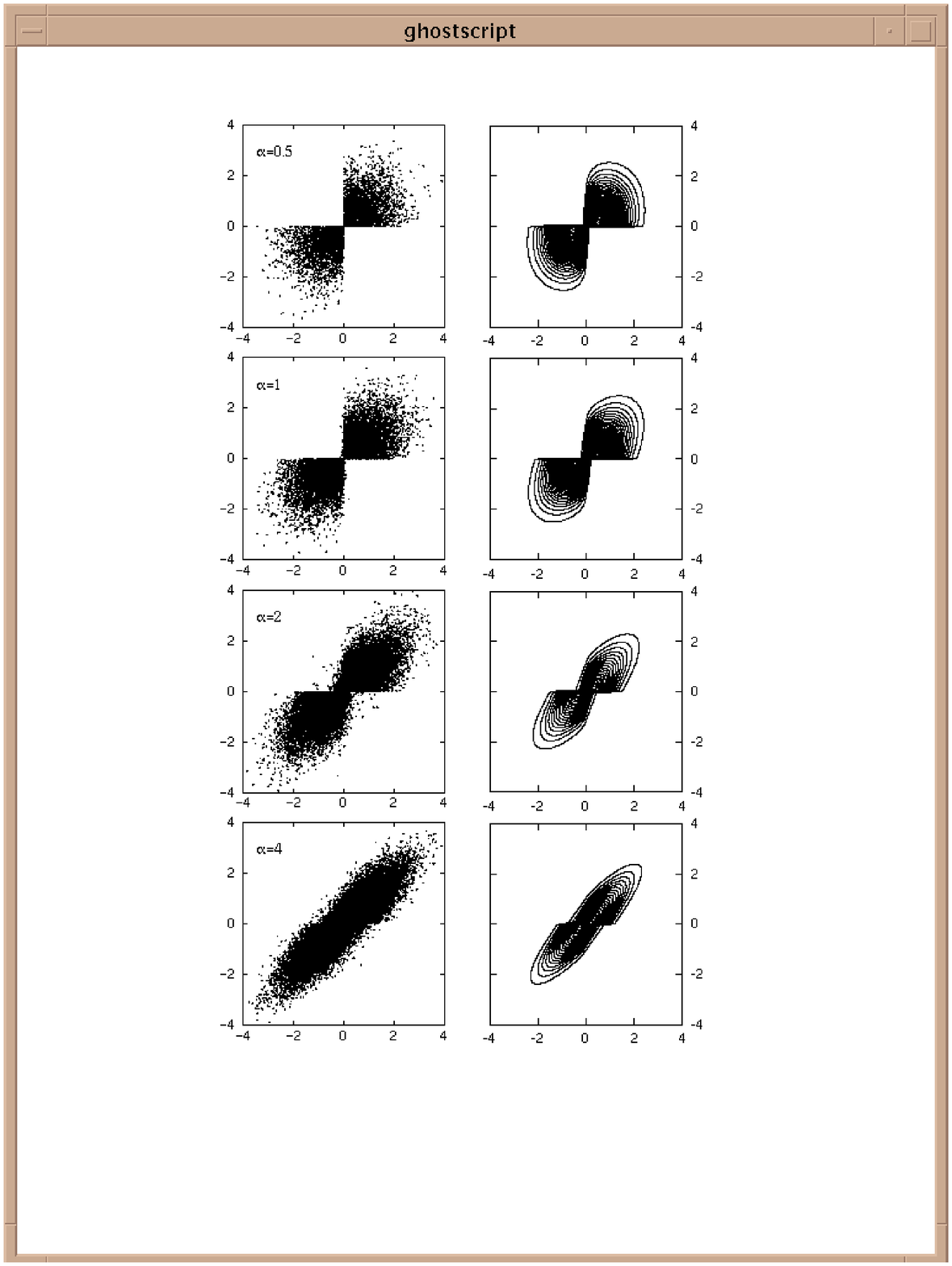}\hspace*{21mm}}
\vspace*{-45mm}
\caption{Comparison between the large $\alpha$ approximation of the
theory and numerical simulations of on-line Adatron learning with
$N\!=\!10,000$ and $\eta\!=\!2$.
Scatter plots (left): observed student and teacher fields
$(x,y)\!=\!(\bJ\inn\bxi,\bB\inn\bxi)$ as measured at time
$t=10$ during simulations,
for the data in $\set$, drawn in the $(x,y)$ plane. Contour plots  (right): corresponding
predictions for the joint field distribution $P[x,y]$, according to the
approximated theory. Training set sizes:
$\alpha\!=\!0.5,1,2,4$ (from top to bottom).}
\vspace*{-5mm}
\label{fig:approx1d}
\end{figure}
\clearpage

\noindent
rules, as we have
seen earlier. However, numerical solution of the coupled equations
(\ref{eq:approx2_fouraverages},\ref{eq:approx2_onlinedQdtdRdt},\ref{eq:approx2_onlinedPdt})
shows quite 
clearly that for the more sophisticated non-Hebbian rules such as Perceptron
and AdaTron, which are of an error correcting nature (i.e. where where
changes are made only when student and teacher disagree),  the
conditionally-Gaussian
 approximation
is less accurate than the
previously investigated  
large $\alpha$ approximation, in spite of the
fact that the 
latter involved much simpler equations. Apparently the
generally
non-Gaussian nature of the conditional distribution $P[x|y]$, and thereby of the measure $M[x|y]$, is of
crucial importance. It is not good enough to try getting away
with allowing the $y$-dependent averages $\overline{x}(y)$ and
variances $\Delta(y)$ to be non-trivial functions.
With conditionally-Gaussian measures $M[x|y]$ it turns out that
generating the right width of the conditional distributions $P[x|y]$ inevitably
introduces tails for $P[x|y]$ which spill into the $xy<0$ region, which are
found to be absent in error-correcting learning rules such as
Perceptron and Adatron.
This picture is consistent with figures
\ref{fig:approx1c} and \ref{fig:approx1d}, where we can observe that
for any fixed value of the teacher field $y$ the remaining
marginal distribution for $x$ is generally not symmetric around
its ($y$-dependent) average. We conclude that the 
conditionally-Gaussian approximation is generally inferior to the 
large $\alpha$ approximation. We
will not waste paper by producing large numbers of graphs to
illustrate this explicitly and comprehensively,
but we will rather draw the conditionally-Gaussian predictions
together with those of the other approximations and of the full
theory, by way of illustration.

\subsection{Partially Annealed Approximation and Full Equations}

\begin{figure}[t]
\centering
\vspace*{116mm}
\hbox to
\hsize{\hspace*{-5mm}\includegraphics{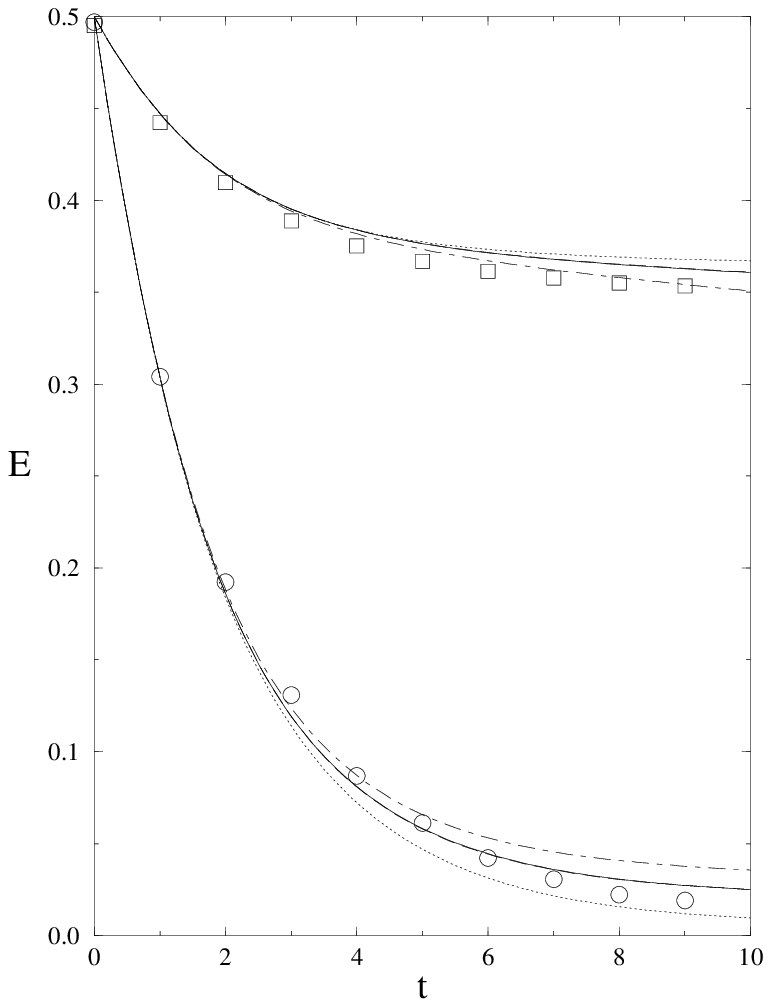}\hspace*{5mm}}
\vspace*{-5mm}
\hbox to
\hsize{\hspace*{80mm}\includegraphics{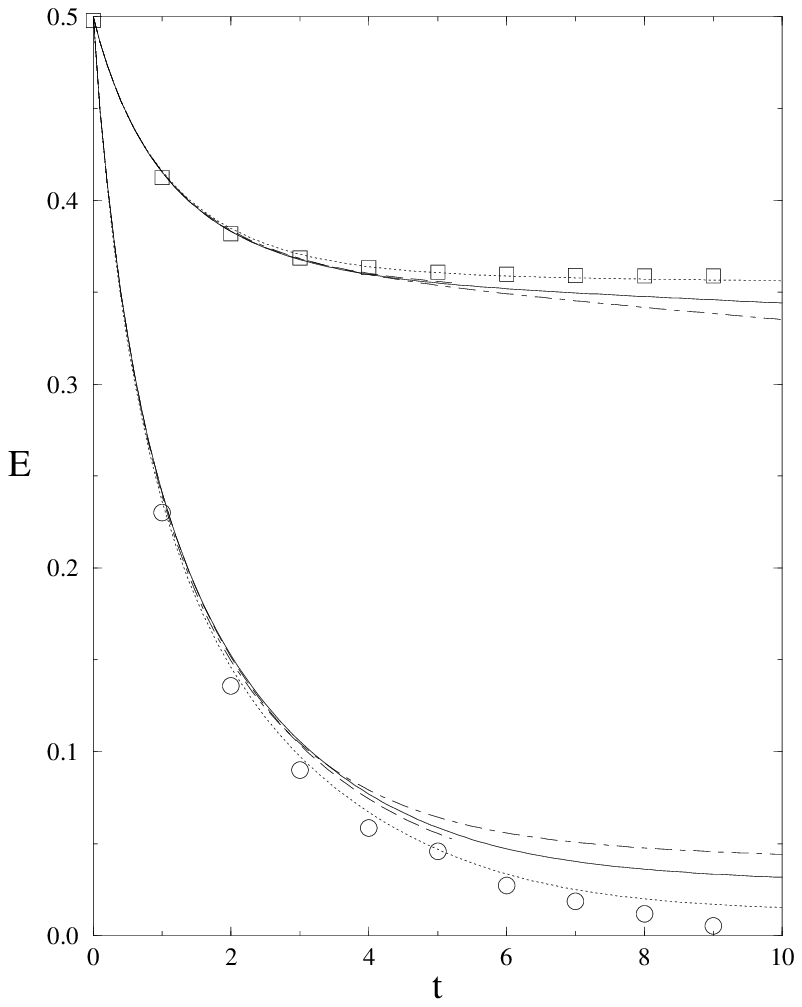}\hspace*{-80mm}}
\vspace*{-15mm}
\caption{Comparison between the full numerical solution of our equations, as well as the three
approximations of the theory, and the results of doing  numerical simulations of on-line learning with
$N=10,000$ and $\alpha=1$.
Markers: training errors $E_{\rm t}$ (circles) and generalisation errors
$E_{\rm g}$
(squares); finite size effects are of the order of the size of te
markers. 
Lines: theoretical predictions for training errors (lower) and
generalisation errors (upper) as functions of time, according to the theory. The different line types refer to: full
equations (solid), annealed approximation (dashed),
conditionally-Gaussian  approximation (dashed-dotted) and large $\alpha$ approximation
(dotted) (note: the dashed and solid curves fall virtually on top of
one another). 
 Left picture: Perceptron learning, with $\eta=\frac{1}{2}$.
Right picture: AdaTron learning, with $\eta=\frac{3}{2}$.}
\label{fig:full}
\end{figure}

The partially annealed approximation and the full theory are both expected
to improve upon the large $\alpha$ approximation (note that
the partially annealed approximation can be seen as an improved version of
the large $\alpha$ approximation, similar in structure but
valid also for small $\alpha$, i.e. large $B$). Although
the partially annealed approximation does not involve a
functional saddle-point equation to be solved (which improves
numerical speed), it shares with the full theory the appearance of
nested (Gaussian) integrals, namely those appearing  in the
function $\Phi[x,y]$ and in the saddle-point equation for $q$.
Thus, solution of both the full theory and of the partially annealed approximation
involves a significant amount of CPU time (avoiding standard instabilities
of discretised diffusion equations sets further limits on the maximum size of the
time discretisation, dependent on the
field resolution \cite{Recipes}), which
implies that we have to reduce our ambition and restrict the number of experiments to
a few typical ones.

We will thus investigate two examples, both with
$\alpha=1$: on-line Perceptron learning with $\eta=\frac{1}{2}$,
and on-line AdaTron learning with $\eta=\frac{3}{2}$. We solve
numerically the full equations of our theory, i.e. the macroscopic dynamical
laws
(\ref{eq:summaryonlinedQdtdRdt},\ref{eq:summaryonlinedPdt}) with
the order parameters calculated at each time step by solving
(\ref{eq:summarysaddle_q},\ref{eq:summarysaddle_M}), and show  in
figure \ref{fig:full}
the training and generalisation errors as functions of time together
with the corresponding values as measured during numerical simulations,
with systems of size $N=10,000$. 
In addition, we plot in the same picture,
for
comparison, the training- and generalisation errors obtained by
numerical solution of the three approximated theories as derived
in the previous section.
In comparing curves we have to take into
account that those describing the large $\alpha$ approximation were 
generated upon solving the diffusion equation with a significantly 
higher numerical field resolution
($\Delta x=0.015$) than the others (where we used $\Delta x=0.05$), 
because of CPU limitations. A restricted field resolution is likely to
be more critical  at large times, where the probability weight in the
$xy<0$ region, responsible for the residual error and for the
non-stationarity of the dynamics, is highly concentrated close to the
decision boundary $x=0$. 
Especially for large times, 
we should therefore expect the full theory, the
conditionally-Gaussian approximation, and the partially annealed
approximation to all three perform better in reality 
than what is suggested by the numerical solutions of their equations
as shown in figure \ref{fig:full}. 
This is particularly true for AdaTron
learning, where even for $\eta>1$ (where we do not expect to observe a
$\delta$-singularity) the field distributions still tend to
develop a jump discontinuity at $x=0$. 
It turns out that the curves of the full theory and those of the
partially annealed approximation are very close (virtually on top of
one another for the case of Perceptron learning) 
in figure \ref{fig:full}; apparently for the learning times considered
here there is no real need to evaluate the full theory. 

Finally, we show in figure \ref{fig:fullfields} for both the full theory and
for the simulation experiments  the two distributions
$P^\pm(x)=\int\!dy ~P[x,y]\theta[\pm y]$ for the student fields,
given a specified sign of the teacher field $y$  (and thus a given teacher
output), corresponding to the same experiments. Note that
$P(x)=P^+(x)+P^-(x)$. 
The pictures in figure  \ref{fig:fullfields} 
again illustrate quite clearly the difference between learning
with restricted training sets and learning with infinite training
sets: in the former case the desired agreement $xy>0$ between student and
teacher is achieved by a qualitative {\em deformation} of $P[x|y]$, away from
the initial Gaussian shape, rather than by adaptation of the first
and second order
moments. 
\clearpage

\begin{figure}[h]
\centering
\vspace*{80mm}
\hbox to
\hsize{\hspace*{3mm}\includegraphics{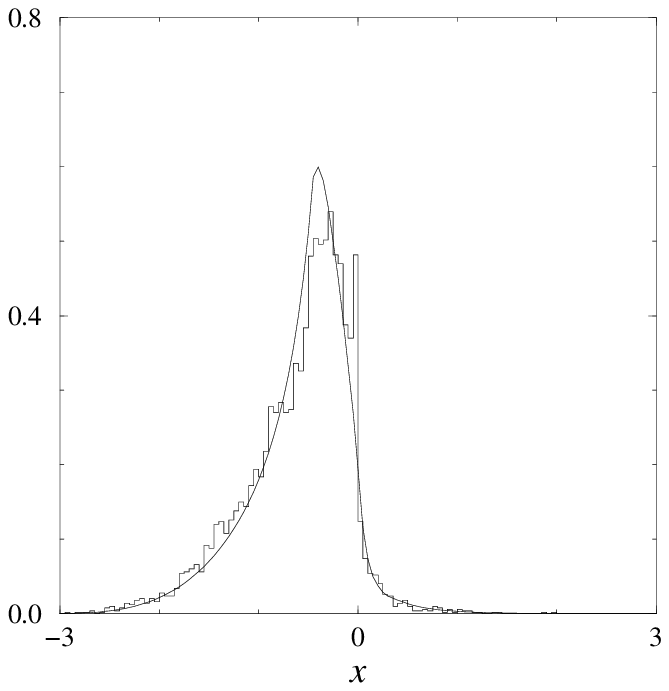}\hspace*{-3mm}}
\vspace*{-5mm}
\hbox to
\hsize{\hspace*{72mm}\includegraphics{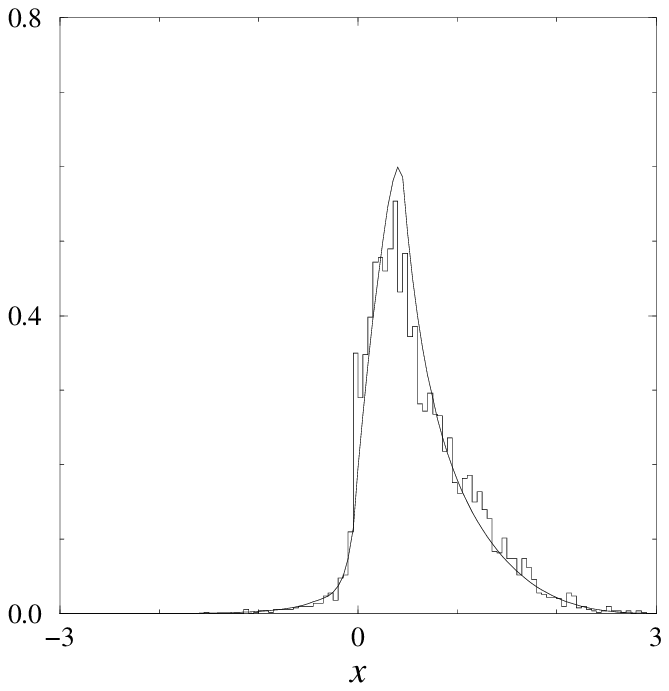}\hspace*{-72mm}}
\vspace*{65mm}
\hbox to
\hsize{\hspace*{3mm}\includegraphics{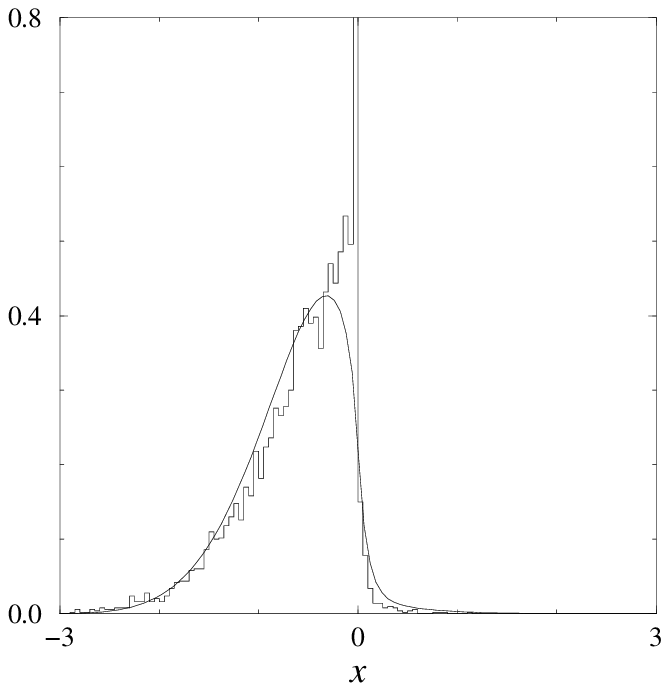}\hspace*{-3mm}}
\vspace*{-5mm}
\hbox to
\hsize{\hspace*{72mm}\includegraphics{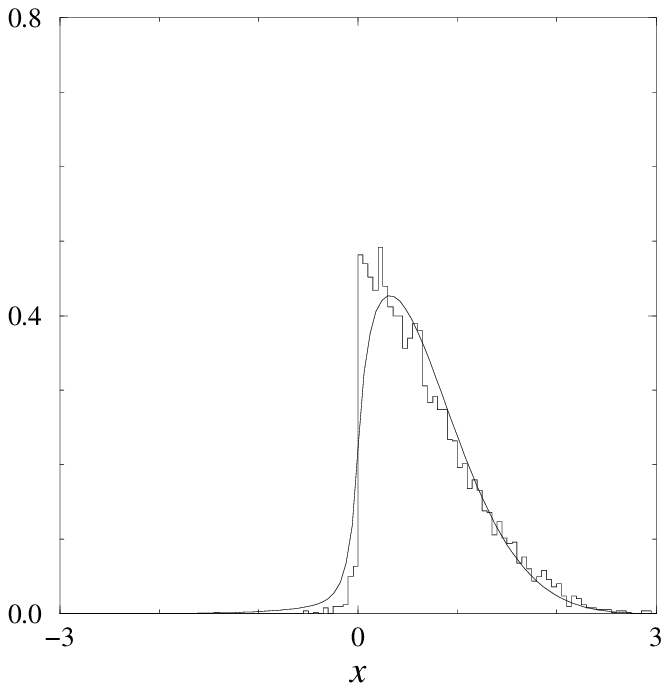}\hspace*{-72mm}}
\vspace*{-15mm}
\caption{Comparison between the full numerical solution of our equations and the results of doing
 numerical simulations of on-line learning with
$N=10,000$ and $\alpha=1$.
Histograms: conditional student field distributions $P^\pm(x)=\int\!dy P[x,y]\theta[\pm y]$ as measured at time $t=5$.
Smooth curves: corresponding theoretical predictions.
Upper pictures: Perceptron learning, with $\eta=1$ (left: $P^-(x)$, right: $P^+(x)$).
Lower pictures: AdaTron learning, with $\eta=\frac{3}{2}$ (left: $P^-(x)$, right: $P^+(x)$).}
\label{fig:fullfields}
\end{figure}

Our restricted resolution numerics obviously have 
difficulty in reproducing the discontinuous behaviour of $P^\pm(x)$ near $x=0$ for
on-line Adatron learning (as expected), which explains why in this regime the
simplest large $\alpha$ approximation (which can be numerically evaluated with almost
arbitrarily high field resolution) appears to outperform the more
sophisticated versions of the theory (which CPU limitations force us
to evaluate with rather limited field resolution), according to figure
\ref{fig:full}.

We conclude from the results in this section that our full theory indeed gives
an adequate description of the macroscopic process, and that the
partially annealed approximation is almost equivalent in performance
to the full theory. 
 As mentioned
before, the conditionally-Gaussian approximation 
performs generally poorly (except,
as we have seen earlier, for the simple Hebian rule). 
Which of the remaining three versions of our theory to use in practice
will 
clearly depend on the accuracy
constraints and available CPU time of the user, with the full
theory at the higher end of the market (in principle very 
accurate, but almost too CPU expensive to work out and exploit
properly), with the large $\alpha$
approximation on the lower end (reasonably acurate, but very cheap), and with the
annealed approximation as a sensible compromise in between these two.

\section{Discussion}

Our aim in this sequel paper was to work out the general theory developed in
\cite{CoolenSaad1} for several supervised (on-line and batch, linear
and non-linear) learning 
scenarios in single-layer perceptrons, to develop a number of
systematic approximations from the 
full set of equations, and to test the theory and its approximations  against both exactly
solvable benchmarks and extensive numerical simulations.
The theory,
built on the dynamical replica formalism \cite{Coolenetal}, was designed to predict the evolution of training- and generalisation errors,
via a non-linear diffusion equation for the joint distribution of student and teacher fields, in
the regime where the size $p$ of the (randomly composed) training set
scales as
$p=\alpha N$, with $0<\alpha<\infty$ and where $N$ denotes the number of inputs. In this regime the input data will in due course be
recycled, as a result of which complicated correlations develop
between the student weights and the realisations of the data
vectors (with their corresponding teacher answers) in the training set;
 the student fields are no longer described by Gaussian
distributions, training- and generalisation errors will no longer be identical,
and the more traditional and familiar statistical mechanical
formalism as developed for infinite training sets consequently breaks down.

We have first worked out our equations explicitly for the
special
case of Hebbian learning, where the availability of exact results,
derived directly from the microscopic equations, allows us to perform
a critical test of the theory.
For batch Hebbian learning we can demonstrate explicitly that our
theory is fully exact.
For on-line Hebbian learning, on the other hand,
proving or disproving full exactness requires
solving a non-trivial functional saddle-point equation analytically,
which we have not yet been able to do. Nevertheless, we can prove that
our theory is exact (i) with respect to its predictions for $Q$, $R$
and $E_{\rm g}$, (ii) with respect to the first moments of the
conditional field distributions $P[x|y]$ (for any $y\in\Re$), and
(iii) in the stationary
state. In order to also generate predictions for intermediate times
we have constructed an approximate
solution of our equations, which is found to
describe the results of performing numerical simulations of on-line
Hebbian learning essentially perfectly.

No exact
benchmark solution is available for non-Hebbian (i.e. non-trivial)
learning rules, leaving numerical simulations as
the only yardstick against which to test our theory.
Motivated by the need 
to solve a functional
saddle-point equation at each time step in the full theory, 
and by the presence of nested integrations,
we have constructed a number of systematic approximations to the
original equations.
We have compared the predictions of the full theory and of the three approximation
schemes with one another and with the results obtained upon performing 
numerical simulations of non-linear learning rules, such as Perceptron and
AdaTron, in large perceptrons (of size $N=10,000$), with various
values of learning rates $\eta$ and relative training set sizes
$\alpha$.
One of the approximations, a conditionally-Gaussian
saddle-point approximation in the spirit of the particular approximation that
was found to work perfectly for Hebbian learning, turned out to 
perform badly for general non-Hebbian rules.
The other two approximations, the large $\alpha$ approximation and the
partially annealed approximation, each have their specific usefulness; the
former is extremely simple and fast, whereas the latter is
overall more accurate, but more expensive in its CPU requirements (so
that in practice its true accurateness cannot always be realised).
Yet, the large $\alpha$ approximation still works 
remarkably well, even for small $\alpha$, in spite of 
it being so simple that
it can be written as a fully explicit set of equations for $Q$, $R$ and the joint field
distribution  $P[x,y]$ only. For on-line learning these equations can be
simplified to
\bd
\frac{d}{dt}Q=2\eta \bra x \G[x,y]\ket +\eta^2 \bra \G^2[x,y]\ket
~~~~~~~~~~~~~~~~
\frac{d}{dt}R =
\eta \bra y \G[x,y]\ket
\ed
\bd
\frac{d}{dt}P[x|y]=\frac{1}{\alpha}\int\!dx^\prime P[x^\prime|y]\left\{\room
\delta[x\minus
x^\prime\minus \eta\G[x^\prime\!,y]]\minus \delta[x\minus
x^\prime]\right\}
+ \frac{1}{2}\eta^2 
\frac{\partial^2}{\partial x^2}P[x|y]
\bra \G^2[x^\prime\!,y^\prime]\ket
\ed
\bd
- \eta\frac{\partial}{\partial x} \left\{\room
P[x|y]\bigbra \G[x^\prime\!,y^\prime]
\left[yy^\prime +\frac{(x\minus
\overline{x}(y))(\overline{x}(y^\prime)\minus Ry^\prime)+(x\minus
Ry)(x^\prime\minus \overline{x}(y^\prime))}{Q-R^2}\right] \bigket
\right\}
\ed
with $\bra f[x,y]\ket=\int\!Dy dx ~
P[x|y] f[x,y]$ and
$\overline{x}(y)=\int\!dx~xP[x|y]$. 
The observed accurateness of these simple equations in the small $\alpha$
regime suggests that for $\alpha\to 0$ the leading term in the
diffusion equation for $P[x|y]$ is the first term in the right-hand
side, which reflects the direct effect of pattern recycling, and which
indeed has not  been approximated. 

 For a discussion of further theoretical developments and refinements of the
present dynamical replica formalism we refer to
\cite{CoolenSaad1}.
We believe that our theory offers an efficient tool
with which to analyse and predict the outcome of learning
processes in single-layer networks. In particular, for those who are primarily interested
in the progress and the outcome of learning processes there is no real
need to understand the full details of the derivation in
\cite{CoolenSaad1}; as in this paper, one can simply adopt the
macroscopic laws of \cite{CoolenSaad1} (or one of the two appropriate approximations, to save CPU time)
as a starting point, and just apply them to the
learning rules as hand.
Generalization to multi-layer networks (with a finite number of
hidden nodes) is straightforward, although numerically intensive \cite{CoolenSaad3}.
The case of noisy teachers can also be studied
with an appropriate extension of the present formalism \cite{Mace}, involving a joint distribution
of three rather than two fields (namely those of student, `clean' teacher, and `noisy' teacher).
 In the example applications worked out so
far in this paper (Hebbian learning, Perceptron learning and AdaTron learning)
our formalism has been found to be either exact or an excellent
approximation. It is not realistic to expect that simpler theories can be
found with a similar level of accuracy. 
While putting the finishing touch to this manuscript a preprint was
communicated \cite{Wong} in which the authors apply the cavity method 
to the present problem. They manage to keep their theory relatively
simple by restrict themselves in several serious
ways (to batch learning only, and to gradient descent learning rules in order
to use FDT relations) and by applying their theory only to a linear
learning rule. Here also the present theory would have been both
simpler and  exact.    
An exact theory for both on-line and batch learning and for arbitrary 
learning rules can be
constructed \cite{Heimel} using a suitable adaptation of the path integral methods
as in \cite{Horner}, with the obvious appeal of full mathematical rigour at any time,
but in describing transients it is going to be more complicated than
the present one as it
will be built around macroscopic observables with two time
arguments rather than one (correlation- and response functions)
and will take the form of an effective single weight process with
a dynamics with coloured stochastic noise and retarded
self-interactions.

\subsection*{Acknowledgement}

DS and ACCC gratefully acknowledge support by the Engineering and
Physical Sciences Research Council
(grant GR/L52093) and by the London Mathematical Society (grant 4415).

\end{document}